\documentstyle{elsart}
\newtheorem{theorem}{Theorem}
\newtheorem{proposition}{Proposition}
\newtheorem{corollary}{Corollary}
\newtheorem{lemma}{Lemma}
\def\KP{{\rm KP}}

\def\Max{{\rm Max}}

\begin{document}

\begin{frontmatter}



\def\C{{\mathbb C}}
\def\Max{{\rm Max}}
\def\err{{\rm err}}
\def\det{{\rm det}}
\def\Pr{{\rm Pr}}
\def\sign{{\rm sign}}
\def\bin{{\rm bin}}
\def\str{{\rm str}}
\def\K{{\rm K}}
\def\KP{{\rm KP}}
\def\I{{\rm I}}
\def\T{{\rm Time}}

\title{Degrees of randomized computability: decomposition into atoms}



\author{Vladimir V. V'yugin\thanksref{label1}}
\thanks[label1]{This research was partially supported by Russian foundation
for fundamental research:  20-01-00203-a.
}

\address{Institute for Information Transmission Problems,
Russian Academy of Sciences,
Bol'shoi Karetnyi per. 19, Moscow GSP-4, 127994, Russia.
e-mail vyugin@iitp.ru}

\begin{abstract}
In this paper we study the structural properties of LV-degrees of the algebra of collections
of sequences that are non-negligible in the sense that they can be computed by
a probabilistic algorithm with positive probability.
We construct atoms and infinitely divisible elements of this algebra generated
by sequences, which cannot be Martin-L\"of random and, moreover, these sequences
cannot be Turing equivalent to random sequences. The constructions
are based on the corresponding templates which can be used for defining the special
LV-degrees. In particular, we present the template for defining atoms of the algebra
of LV-degrees and obtain the decomposition of the maximal LV-degree into
a countable sequence of atoms and their non-zero complement -- infinitely
divisible LV-degree. We apply the templates to establish new facts about specific 
LV-degrees, such as the LV-degree of the collection of sequences of hyperimmune degree.
We construct atoms defined by collections of hyperimmune sequences, moreover,
a representation of LV-degree of the collection of all hyperimmune sequences
will be obtained in the form of a union of an infinite sequence of atoms and
an infinitely divisible element.
\end{abstract}

\end{frontmatter}

\maketitle

\section{Introduction}\label{intr-1}

The problem of the existence of nonstochastic objects has been discussed in
seventies at Kolmogorov's seminar in the Moscow State University (see also
Cover et al.~\cite{CGG89}). Following Levin~\cite{Lev84}, Levin and V'yugin~\cite{LeV77},
V'yugin~\cite{Vyu76, Vyu82, Vyu2012}),
the most suitable objects for such a study are infinite binary sequences and
the problem should be considered in the information aspect, that is, the problem is
whether we can generate ``nonstochastic information'' using a probabilistic
Turing machine and what types of nonstochastic information can be generated
using probabilistic these machines.

We will study the properties (collections or Borel sets) of infinite binary sequences,
as carriers of certain information, i.e. such properties should
be invariant with respect to the encoding methods. We will consider
the encoding methods of the most general type -- algorithmic operators.

The algebra of collections (of infinite sequences) that are closed under Turing equivalence
were introduced by Levin and V'yugin~\cite{LeV77} and studied by V'yugin~\cite{Vyu82}.
Roughly speaking, given two such collections $A$ and $B$, $A\preceq B$ in this ordering
if $A\setminus B$ is negligible, i.e. the Levin's a priory (semi)measure of this set
is equal to zero. This is equivalent to the fact that no probabilistic Turing machine
can produce a sequence from this set with a positive probability. See about the
a priory semimeasure in Zvonkin and Levin~\cite{ZvL70}, Levin and V'yugin~\cite{LeV77}, 
and in Solomonoff~\cite{Sol64, Sol64a}.

We call any two invariant collections $A$ and $B$
equivalent, $A\sim B$, if $A\preceq B$ and $B\preceq A$, i.e., these sets differ
on a negligible set. We consider the factor algebra with respect to this equivalence,
its element is defined as $[A]=\{B:B\sim A\}$.

The degree structure associated with this ordering is a Boolean algebra which
was called by V'yugin~\cite{Vyu82} the algebra of invariant properties.
This algebra was recently called by Bienvenu and Patey~\cite{BiP2017} and in
Holzl and Porter~\cite{HoP2021} the algebra of Levin-V'yugin degrees (or LV-degrees).
The idea of studying this algebra was put forward by Leonid Levin.

In this paper we study structural properties of LV-degrees of collections
of sequences that are non-negligible in the sense that they can be computed by
a probabilistic algorithm with positive probability. The equivalence class
of all negligible collections defines the zero element while
all infinite sequences defines the maximal -- unit element, of this algebra.

Two natural elements of this algebra can be distinguished. This is the element generated
by non-computable Martin-L\"of random sequences and the element generated by computable
sequences.

It was proved by V'yugin~\cite{Vyu76, Vyu82} that the complement
of these two elements is non-zero. Using the probabilistic Turing machines, we can generate
infinite sequences which are non-random with respect to any computable measure
and, moreover, they cannot be Turing equivalent to random sequences. We say that
such sequences carry non-random information.

In this work, we will study the properties of of LV-algebra. Levin and V'yugin~\cite{LeV77} 
pointed out that all non-computable Martin-L\"of sequences define
the atom of LV-algebra in the sense that such an element cannot be represented as a union
of two incomparable non-zero elements. Computable sequences form the atom in a trivial way.

In this paper we construct infinitely many other atoms defined by collections of non-random
sequences. We will also show that the complement of all atoms of the LV-algebra is
nontrivial. The complement of all atoms is an infinitely divisible element.
Thus, we get a representation of the unit element of LV-degrees in the form of
a union of an infinite sequence of atoms, two of which have the natural interpretation,
and an infinitely divisible element. The constructions are based on the corresponding
templates. In particular, we present the template for defining atoms of the algebra
of LV-degrees.

Also, we correct and improve the construction of the atoms from V'yugin~\cite{Vyu82},
which is insufficient to achieve the desired result. The author is grateful
to Rupert Holzl and Cristopher Porter who noticed this insufficiency.

An excellent analysis of the relationship between LV-degrees and Turing degrees is
given in the review by Holzl and Porter~\cite{HoP2021}, where the idea of the
construction from V'yugin~\cite{Vyu82} is also explained.

We will point out the connection between the properties of LV-algebra
and classical properties from computability theory. In particular,
we apply results on the interactions between notions of randomness and Turing
reducibility to establish new facts about specific LV-degrees, such as
the LV-degrees of the collections of hyperimmune sequences
which are characteristic sequences of hyperimmune sets.
We will construct atoms and infinitely divisible element defined by collections
of hyperimmune sequences. Thus, a representation
of LV-degree of the collection of all hyperimmune sequences will be obtained in the
form of a union of an infinite sequence of atoms and an infinitely divisible element.

\section{Preliminaries}\label{prelimin-1}

Let $\Xi$ be the set of all finite binary sequences,
$\Omega$ be the set of all infinite binary sequences, $\lambda$ be the empty sequence.
In what follows by a sequence (finite or infinite) we mean the binary sequence, i.e.,
the sequence $\omega_1\omega_2\dots$, where $\omega_i\in\{0,1\}$ for $i=1,2,\dots$.
For any finite or infinite $\omega=\omega_1\dots\omega_n\dots$, we
denote its prefix (initial fragment) of length $n$ as $\omega^n=\omega_1\dots\omega_n$.
We write $x\subseteq y$ if a sequence $y$ is an extension
of a sequence $x$, $l(x)$ is the length of $x$.

Let $\cal R$ be the set of all real numbers extended by adding
the infinities $-\infty$ and $+\infty$, $\cal N$ and $\cal Q$ -- be the sets
of all positive integer numbers and of all rational numbers.  Let $[r]$ denotes
the integer part of a real number $r$.

We assume that the reader is familiar with the basics of computability
and algorithmic randomness theory
(for instance, the material covered in Rogers~\cite{Rog67}, Soare~\cite{Soa2016},
Nies~\cite{Nie2009}, Downey and Hirschfeldt~\cite{DH2010}, G{\'a}cs~\cite{Gac12},
Vereshchagin et al.~\cite{She2007}, and Li and Vitanyi~\cite{LiV97}).
                          
We need a one-to-one enumeration of all ordered pairs of positive integers.
We fix some form of this enumeration. We use the natural correspondence between
finite binary sequences and nonnegative integers:
$\lambda$--$0$, $0$--$1$, $1$--$2$, $00$--$3$,
$01$--$4$, $11$--$5$, $000$--$6,\dots$.
We will identify any ordered pair of positive integer numbers $\langle i,j\rangle$
and its ordinal number.
A one-to-one enumeration of all ordered triples $\langle i,j,k\rangle$ also can be defined
in a similar way: $\langle i,j,k\rangle=\langle \langle i,j\rangle,k\rangle$.
We denote the inverse functions $[\langle i,j\rangle]_1=i$ and $[\langle i,j\rangle]_2=j$.
Also, $[\langle i,j,k\rangle]_1=i$, $[\langle i,j,k\rangle]_2=j$ and
$[\langle i,j,k\rangle]_3=k$.

We fix a model of computation. Algorithms may be regarded as
Turing machines and so the notion of a program and a time of computation
will be well-defined. Our considerations will be invariant under polynomial
computation time, so the results will be machine-independent.
An algorithm transforms finite objects into finite objects.
Integer and rational numbers (but no reals) are examples of finite objects.
Finite sequences of finite objects are again finite objects.
The main property of finite objects that we use is that
they can be enumerated with positive integers, and therefore they can be
arguments of computable (partial recursive) functions and algorithms.

A function $f$ is called partial recursive if there is an algorithm
(Turing machine) computing values of $f$. For any input $x$
the corresponding Turing machine when fed with $x$ stops after several steps
and outputs the result $f(x)$ if $f(x)$ is defined and never stops otherwise.
We call a function $f$ total if $f(x)$ is defined for every $x$.

Define
\[
  f^n(x)=
  \left\{
    \begin{array}{l}
      f(x) \mbox{ if } f(x)\mbox{ is computed in }n\mbox{ steps}
    \\
      \infty\mbox{, otherwise},
    \end{array}
  \right.
\]
where $f$ is a partial recursive function and $n$ is a positive integer number.

We will use the recursive sequence $\phi_i$ of all partial recursive functions.
This means that there is partial recursive universal
function $U(i,x)=\phi(x)$ such that for any partial recursive function $f$ there exists
an $i$ such that $\phi_i(x)=U(i,x)=f(x)$ for each $x$, where both sides if this
equality are defined or undefined simultaneously.

A set of finite objects is called recursively enumerable if it is a domain 
of some computable function. A nonempty set $A$ is recursively enumerable 
if and only if it is the range of some total recursive
(computable) function. A set $A$ of finite objects is
called (algorithmically) decidable (recursive) if $A$ and the complement of $A$ are
recursively enumerable.

Let $A$ be a set of all finite objects of certain type.
A function $f\colon A\rightarrow\cal R$ is called lower semicomputable if
$\{(r,x): x\in\Xi,\ r\in{\cal Q}, r<f(x)\}$ is a recursively
enumerable set. This means that there is an algorithm which when fed with
a rational number $r$ and a finite object $x$ eventually stops if
$r<f(x)$ and never stops otherwise. In other words, the semicomputability
of $f$ means that if $f(x)>r$ this fact will sooner or later be learned,
whereas if $f(x)\leq r$ we may be for ever uncertain.

We use also a concept of computable operator (operation) on $\Xi\bigcup\Omega$
(Zvonkin and Levin~\cite{ZvL70}, Uspenskyi et al.~\cite{USS90}).
Let $\hat F$ be a recursively enumerable set of
ordered pairs of finite sequences (graph of a computable operator) satisfying
the following properties:
\begin{itemize}
\item{}
$(x,\lambda)\in\hat F$ for any $x$, where $\lambda$ is the empty sequence;
\item{}
if $(x,y)\in\hat F$, $x\subseteq x'$ and $y'\subseteq y$ then $(x',y')\in\hat F$;
\item{}
if $(x,y)\in\hat F$ and $(x,y')\in\hat F$ then $y\subseteq y'$ or $y'\subseteq y$.
\end{itemize}

A computable operator $F$ is defined using the graph $\hat F$ as follows
$$
F(\omega)=\sup\{y: \exists x(x\subseteq\omega\&(x,y)\in\hat F\},
$$
where $\omega\in\Omega\bigcup\Xi$ and $\sup$ is under partial ordering
$\subseteq$ on $\Xi$.

Informally, the computable operator $F$ is defined by some algorithm
which when fed with an infinite or a finite sequence $\omega$ takes it
sequentially bit by bit, processes it and produces an output sequence
also sequentially bit by bit.

In what follows we will use the modified version of computable operator, where
\begin{eqnarray}
\tilde F(x)=\sup\{y: l(y)\le l(x)\&\exists x'(x'\subseteq x\&(x',y)\in\hat F^{l(x)}\}
\label{modify-operator-1}
\end{eqnarray}
for any $x\in\Xi$, where $\sup$ is taken under partial ordering
$x\subseteq y$ and $F^{l(x)}$ is the finite subset of elements of $\hat F$
enumerated in $l(x)$ steps. Let $(x,\lambda)\in\hat F^0$ for every $x$.
By definition for modified operator, $l(\tilde F(x))\le l(x)$ for each finite 
sequence $x$ and $\tilde F(\omega)=F(\omega)$ for each infinite $\omega$,
where $\tilde F(\omega)=\sup_n \tilde F(\omega^n)$. 
For any finite sequence $x$ the value of $\tilde F(x)$ is defined in $l(x)$ steps 
of computation.

We will use the uniformly computable sequence of all computable operators
$\{F_i\}$ such that given $i$ and $\omega$ some algorithm computes the
value $F_i(\omega)$ and for any computable operator $F$ there exists an $i$ such that
$F(\omega)=F_i(\omega)$ for each $\omega$. This sequence is defined by a recursively
enumerable set $\cal F$ of triples $(i,x,y)$ such that for any $i$ the set
$\hat F=\{(x,y):(i,x,y)\in {\cal F}\}$ defines the computable operator $F_i$
and for each computable operator $F$ an $i$ exists such that $F=F_i$.
We transform this sequence to a sequence $\{\tilde F_i\}$
as it was done by (\ref{modify-operator-1}).

A real-valued non-negative function $P:\Xi\to{\cal R}$ is called semimeasure if
\begin{eqnarray}
P(\lambda)\le 1,
\nonumber
\\
P(x0)+P(x1)\le P(x)
\label{semi-1}
\end{eqnarray}
for all $x\in\Xi$. We will consider lower semicomputable
semimeasures $P$; this means that the set
$\{(r,x):r\in {\cal Q},\ r<P(x)\}$ is recursively enumerable.

Solomonoff~\cite{Sol64, Sol64a} proposed ideas for defining the a priori probability
distribution on the basis of the general theory of algorithms.
Levin in~\cite{ZvL70} and~\cite{LeV77} gives a precise
form of Solomonoff's ideas in a concept of a maximal lower semicomputable semimeasure
(see also Li and Vytanyi~\cite{LiV97}, Section 4.5.)
Levin proved that there exists the maximal to within a multiplicative
positive constant factor semimeasure $M$ semicomputable from below, i.e.
for every semimeasure $P$ semicomputable from below a positive constant $c$
exists such that the inequality
\begin{equation}\label{M-ineq}
cM(x)\ge P(x)
\end{equation}
holds for all $x$. The semimeasure $M$ is called a priory or universal semimeasure.

A function $P$ is a measure if (\ref{semi-1}) holds, where
both inequality signs $\le$ are replaced with the equality $=$. Any function $P$ satisfying
(\ref{semi-1}) (with equalities) can be extended on all Borel subsets
of $\Omega$. Consider intervals $\Gamma_x=\{\omega\in\Omega:x\subseteq\omega\}$
in $\Omega$ for all $x\in\Xi$. The measure of any such interval can be defined as
$P(\Gamma_x)=P(x)$ and can be extended to all Borel subsets of $\Omega$.

A measure $P$ is computable if there is an algorithm which for any $x\in\Xi$ 
outputs a rational approximation of the real number $P(x)$ with 
a given degree of precision. A typical example of computable measure on $\Omega$ 
is the uniform measure $L$, where $L(\Gamma_x)=2^{-l(x)}$ for every $x\in\Xi$.

For technical reason, for any semimeasure $P$, we consider the maximal measure
$\bar P$ such that $\bar P\le P$. This measure can be defined as
\begin{eqnarray*}
\bar P(x)=\lim\limits_{n\to\infty}\sum\limits_{l(y)=n,x\subseteq y}P(y)
\end{eqnarray*}
(see Levin and V'yugin~\cite{LeV77}).
In general, the measure $\bar P$ is non-computable (and it is not a probability measure,
since $\bar P(\Omega)<1$) even then $P$ is a lower semicomputable semimeasure.
From (\ref{M-ineq}) the inequality $c\bar M(A)\ge\bar P(A)$ follows for each lower
semicomputable semimeasure $P$ and for every Borel set $A$, where $c$ is a positive
constant (the same as in (\ref{M-ineq})). In particular, the measure $\bar P$ is
absolutely continuous with respect to the measure $\bar M$.

Following Levin~\cite{ZvL70, LeV77, Lev84} (see also V'yugin~\cite{Vyu82, Vyu2012}),
the combinations of probabilistic and deterministic processes is considered as
the most general class of processes for generating data. Any probabilistic process
is defined by some computable probability distribution.
Any deterministic process is realized by means of an algorithm.
Algorithmic processes transform sequences generated by
probabilistic processes into new sequences. More precise, a probabilistic
computer is a pair $(P,F)$, where $P$ is a computable probability distribution
(for example, $P=L$) and $F$ is a Turing machine supplied with an additional input tape.
In the process of computation this machine reads on this tape
a sequence $\omega$ distributed according to $P$ and produces a sequence
$\omega'=F(\omega)$ (A correct definition see in~\cite{ZvL70, USS90, Vyu82, LiV97}).
So, we can compute the probability
$$
Q(x)=P\{\omega: x\subseteq F(\omega)\}
$$
of that the result $F(\omega)$ of the computation begins with a
finite sequence $x$. It is easy to see that $Q(x)$ is the lower semicomputable
semimeasure.
Generally, the semimeasure $Q$ can be not a probability distribution in $\Omega$,
since $F(\omega)$ may be finite for some infinite $\omega$.

The converse result is proved in Zvonkin and Levin~\cite{ZvL70}: for every
lower semicomputable semimeasure $Q(x)$ a probabilistic algorithm $(L,F)$ exists such that
$$
Q(x)=L\{\omega\vert x\subseteq F(\omega)\},
$$
for all $x$, where $L(x)=2^{-l(x)}$ is the uniform probability
distribution in the set of all binary sequences.

Therefore, by (\ref{M-ineq}) $M(x)$ defines an asymptotic universal upper bound of
the probability of generating $x$ by probabilistic algorithms.

A set of infinite sequences $U\subseteq\Omega$ is called open if it can be
represented as a union of a sequence of intervals $U=\cup\Gamma_{x_i}$, where $x_i\in\Xi$
for $i=1,2,\dots$. An open set $U$ is effectively open if the function $f(i)=x_i$ is total
computable.

Let $P$ be a computable measure on $\Omega$.
By  Martin-L\"of~\cite{Mar66} test of randomness with respect to the measure $P$
we mean a uniformly recursively enumerable
sequence $\{U_i\}$ of effectively open sets (i.e., $U_i=\cup_j\Gamma_{x_{i,j}}$
for each $i$ and the function $f(i,j)=x_{i,j}$ is computable) such that
$P(U_i)\le 2^{-i}$ for all $i$. The null-set of any test is $\cap_i U_i$.
From definition $P(\cap_i U_i)=0$. An infinite sequence $\omega\in\Omega$ is
called Martin-L\"of random with respect to a computable measure $P$ if
$\omega\not\in\cap_i U_i$ for each test $\{U_i\}$ of randomness
with respect to $P$.

An equvalent definition of Martin-L\"of randomness can be given in terms
of the a priory semimeasure. 
An infinite sequence $\omega$ is Martin-L\"of random with respect to a computable
measure $P$ if and only if a positive constant $c$ exists such that
\begin{equation}
\frac{P(\omega^n)}{M(\omega^n)}\ge c>0
\label{qrit-random-1}
\end{equation}
for every $n$ (see Zvonkin and Levin~\cite{ZvL70},~\cite{LiV97}).

\section{Algebra of LV-degrees}\label{algebra-1}

An infinite sequence $\alpha\in\Omega$ is Turing (or algorithmically) reducible to
an infinite sequence $\beta\in\Omega$ if $\alpha=F(\beta)$ for some
algorithmic operator $F$. Denote this as $\alpha\le_T\beta$.
Two infinite sequences $\alpha$ and $\beta$
are Turing equivalent: $\alpha\equiv_T\beta$, if each of them
reducible to another one. The classes of equivalent sequences
form Turing degrees.

A Borel set $A\subseteq\Omega$ is called
algorithmically (or Turing) invariant if it is together with each
sequence also contains all algorithmically equivalent
sequences. In other words, the set $A$ can be represented as the union
of Turing degrees. For any set $A\subseteq\Omega$, let
$\bar A=\{\omega:\exists\alpha(\alpha\in A\&\alpha\equiv_T\omega)\}$
be the algorithmic closure the set $A$.

Martin-L\"of random sequences should serve as mathematical analogs of sequences
which can be obtained in stochastic processes.
On the other hand, some infinite Martin-L\"of random sequences can be defined
using exact mathematical constructions, that contradicts to our intuition.
For example, the binary representation of the Chaitin number $\sum_n 2^{-\KP(n)}$,
where $\KP(n)$ is the prefix Kolmogorov complexity of a positive integer number $n$,
is Martin-L\"of random with respect to uniform measure on $\Omega$ (see~\cite{USS90}),
another example of the exact mathematical construction of Martin-L\"of random sequence
can be found in Zvonkin and Levin~\cite{ZvL70}.
These examples show that some correction is needed in the interpretation of
the concept of a random sequence.

By de Leeuw--Moore--Shannon--Shapiro theorem $\bar M(\{\alpha\}) = 0$,
where $\alpha$ is any non-computable sequence (Leeuw et al.~\cite{LMSS56}, 
see also Sacks~\cite{Sac63}).
In particular, the Chaitin number, or rather, its binary representation
$\alpha$, cannot be output (with a positive probability) of any probabilistic algorithm.
Similarly, any individual random sequence $\alpha$, defined by a mathematical construction,
cannot be obtained as output in any combination of random and algorithmic processes.

Let any property ${\cal A}$ defines a Borel set $A=\{\omega\in\Omega:{\cal A}(\omega)\}$
such that $\bar M(A)=0$. Then for any probabilistic machine $(L,F)$,
the probability $P(A)=L\{\omega: F(\omega)\in A\}$ of generation a sequence
from $A$ is equal 0. We call such sets {\it negligible}. A set $A$ is negligible
if and only if $L(F^{-1}(A))=0$ for each computable operator $F$, where
$F^{-1}(A)=\{\omega\in\Omega:F(\omega)\in A\}$ (see~\cite{LeV77, Vyu82}).

We consider algorithmic transformations of infinite
sequences, which can be carried out using probabilistic algorithms.
By definition any infinite sequence from the negligible set cannot be obtained
(with positive probability) in any combination of stochastic and algorithmic
processes. 

For example, for any non-computable infinite sequence $\alpha$ the set
$$
\{\omega\in\Omega:\exists F(F(\omega)=\alpha)\}
$$
is negligible.

Let ${\cal B}$ be the Boolean algebra of all algorithmically invariant
Borel subsets (collections of infinite binary sequences) of $\Omega$.
We identify any two sets from ${\cal B}$
which differ by a negligible set. More correctly,
let us consider the equivalence relation on ${\cal B}$
$$
A\sim B \Longleftrightarrow \bar M((A\setminus B)\cup (B\setminus A))=0.
$$
Let $\Upsilon$ be the factor algebra of ${\cal B}$ by the equivalence relation
$\sim$. Denote the equivalence class of any set $A$ by ${\bf a}=[A]$.
The elements of $\Upsilon$ will be called degrees of randomized computability
or LV-degrees.\footnote{V'yugin~\cite{Vyu82} called $\Upsilon$ the algebra 
of invariant properties, this algebra was recently called by 
Bienvenu and Patey~\cite{BiP2017} and in Holzl and Porter~\cite{HoP2021} 
the algebra of Levin-V'yugin degrees (or LV-degrees).}

Define, for any lower semicomputable semimeasure $P$,
$\bar P({\bf a})=\bar P(A)$. Define the Boolean operations on $\Upsilon$:
${\bf a}\cup {\bf b}=[A\cup B]$ and ${\bf a}\cap {\bf b}=[A\cap B]$,
where ${\bf a}=[A]$ and ${\bf b}=[B]$. The partial ordering on $\Upsilon$
is defined:
$$
{\bf a}\preceq {\bf b}\Longleftrightarrow \bar M(A\setminus B)=0.
$$
In what follows, we call {\it standard} any sequence which is algorithmically
equivalent to a sequence Martin-L\"of random with respect to
some computable measure. By definition, any computable
measure of the set of all standard sequences is equal to 1.

Zvonkin and Levin~\cite{ZvL70} (Theorem~3.1) proved that any sequence $\omega$
Martin-L\"of random with respect to a computable measure is computable or 
algorithmically equivalent to a sequence which is Martin-L\"of random with 
respect to the uniform measure.

Therefore, the elements ${\bf r}=[\bar R]$ and ${\bf c}=[C]$ arise naturally,
where $R$ be the set of all sequences Martin-L\"of random with respect
to uniform measure, $\bar R$ be its algorithmic closure. In particular, the set $\bar R$
contains all non-computable sequences random with respect to computable measures
(and all sequences Turing equivalent to such sequences),
Let $C$ be the set of all computable sequences.
Evidently, $\bar M({\bf r})>0$ and $\bar M({\bf c})>0$.

The zero element {\bf 0} of the algebra $\Upsilon$ is the equivalence class
of the empty set.  It consists of all algorithmically invariant negligible
Borel subsets of $\Omega$, $\bar M({\bf 0})=0$.
The maximal element of $\Upsilon$ is ${\bf 1}=[\Omega]$.

By definition ${\bf d}$ is an atom of $\Upsilon$ if
$\bf d\not =\bf 0$ and it cannot be represented as
${\bf d}={\bf a}\cup{\bf b}$, where ${\bf a}\cap{\bf b}={\bf 0}$,
${\bf a}\not ={\bf 0}$ and ${\bf b}\not ={\bf 0}$.

It was first pointed in Levin and V'yugin~\cite{LeV77} that ${\bf r}$ and $\bf c$
are atoms of $\Upsilon$. The proof of this result (which attributes to Levin)
was first given in V'yugin~\cite{Vyu82}. Holzl and Porter~\cite{HoP2021}
also presented the careful proof of this result.
We present a short proof for completeness of presentation.
\begin{theorem} \label{atom-1}
The element ${\bf r}$ is an atom of $\Upsilon$.
\end{theorem}
{\it Proof.} Assume that ${\bf r}={\bf a}\cup{\bf b}$,
where ${\bf a}\cap{\bf b}={\bf 0}$, ${\bf a}\not ={\bf 0}$ and
${\bf b}\not ={\bf 0}$. Then $\bar R=A\cup B$, where ${\bf a}=[A]$ and
${\bf b}=[B]$, where $A$ and $B$ are the algorithmically invariant sets
of infinite sequences. We can assume without loss of generality that $A\cap B=\emptyset$.
Recall that $R$ is the set of all Martin-L\"of random sequences with respect to
the uniform measure. Let $A'=A\cap R$ and $B'=B\cap R$. Since any sequence $\alpha\in A$
is algorithmically equivalent to some sequence from $A'$ and $\bar M(A)>0$,
$\bar M(A')>0$ follows. Analogously $\bar M(B')>0$.

Let $P$ be a probability measure on $\Omega$ absolutely continuous with respect
to $\bar M$, i.e., $\bar M(X)=0$ implies $P(X)=0$ for each Borel set $X$.

Let $\frac{dP}{d\bar M}(\omega)$ be the Radon--Nicodym derivative of $P$ by
the measure $\bar M$, where $P$ is a probability measure on $\Omega$ absolutely
continuous with respect to the measure $\bar M$. By definition
\begin{equation}
P(X)=\int\limits_{X}\frac{dP}{d\bar M}(\omega)d\bar M
\label{radon-defin-1}
\end{equation}
for each Borel set $X$. In particular, (\ref{radon-defin-1}) holds true for
each computable measure $P$ and for the measure $\bar P$, where $P$ is
a lower semicomputable semimeasure.

\begin{lemma}\label{Rad-Nic}
Let a measure $P$ is absolutely continuous with respect to the measure $\bar M$,
$A\subseteq\Omega$ and $\frac{dP}{d\bar M}(\omega)>0$ for each $\omega\in A$.
Then $P(A)=0$ implies $\bar M(A)=0$.
\end{lemma}
{\it Proof.} By (\ref{radon-defin-1}),
$P(A)=\int\limits_{A}\frac{dP}{d\bar M}(\omega)d\bar M$.
It is easy to prove that if $\frac{dP}{d\bar M}(\omega)>0$ for each $\omega\in A$
and $P(A)=0$ then $\bar M(A)=0$.
$\Box$

\begin{corollary} \label{rad-2}
Let $P$ be a computable measure and $A$ consists of $P$-random sequences.
Then $P(A)=0$ implies $\bar M(A)=0$.
\end{corollary}
{\it Proof.} For any random sequence $\omega$
$$
P(\omega^n)/\bar M(\omega^n)\ge P(\omega^n)/M(\omega^n)\ge c>0
$$
holds for every $n$, where $c$ is a constant depending on $\omega$. Then
$\frac{dP}{d\bar M}(\omega)\not =0$ for each $\omega\in A$. By Lemma~\ref{Rad-Nic}
$\bar M(A)=0$. $\Box$

Let us finish the proof of the theorem. If an infinite sequence
$\omega$ is random with respect to the uniform measure then any sequence $\omega'$,
which differs from it in a finite number of bits, is also random.
Then $\omega,\omega'\in R$. Besides, $\omega\equiv_T\omega'$.

We can choose algorithmically invariant Borel sets $A$ and $B$ such that
any two sequences $\alpha\in A$ and $\beta\in B$ are not algorithmically
equivalent. Let $A'=A\cap R$ and $B'=B\cap R$.
Then $\omega,\omega'\in A'$ or $\omega,\omega'\in B'$,
By Corollary~\ref{Rad-Nic} $\bar M(A')>0$ implies $L(A')>0$. Analogously
$\bar M(B')>0$ implies $L(B')>0$.

We apply the Kolmogorov 0 or 1 law to the sequence $f_1, f_2,\dots$
of random variables, where $f_i(\omega)=\omega_i$ are random variables
defined on the probability space $(\Omega,L)$.
It follows from invariant property of the sets $A'$ and $B'$
that for each $n$ they belong to $\sigma$-algebra generated by random variables
$f_n,f_{n+1},\dots$ and then
they belong to the residual $\sigma$-algebra of $f_1, f_2,\dots$.
By Kolmogorov 0 or 1 law $L(A')=0$ or $L(A')=1$ and the same holds for $B'$.
This is a contradiction, since $A'\cap B'=\emptyset$ and $L(A')>0$,
$L(B')>0$. Therefore, $\bf r$ is an atom of $\Upsilon$.
$\Box$

Evidently, $\bf c$ is also an atom of $\Upsilon$.
It is easy to see that $\bf r$ is the single atom of the uniform measure 1.

The atoms $\bf c$ and $\bf r$ be generated by the standard sequences. A question
arises does ${\bf 1}={\bf c}\cup {\bf r}$?\footnote{This would mean that all 
sequences (information) 
that can be generated using probabilistic algorithms are stochastic or computable.}
We prove in Section~\ref{app-1}
that ${\bf 1}\setminus({\bf c}\cup {\bf r})\not={\bf 0}$ and, moreover,
we prove in Section~\ref{atoms-count-1} that there exists an infinite sequence 
of other atoms.

It is easy to show that the set of all atoms of the algebra $\Upsilon$ is at most
countable. To do this, choose for each atom ${\bf a}=[A]$ a union $D_a$ of a finite number
of intervals such that
$$
\bar M((A\setminus D_a)\cup(D_a\setminus A))<(1/4)\bar M(\bf a).
$$
If $\bf a\not =\bf b$ then $D_a\not = D_b$. The set of all $D_a$ is at most
countable.

We will prove in Section~\ref{atoms-count-1} that the set
of all atoms is countable. Let ${\bf a}_1,{\bf a}_2,{\bf a}_3,\dots$ be all atoms of
$\Upsilon$, where ${\bf a}_1={\bf c}$ and ${\bf a}_2={\bf r}$ and the atoms

$\bf c$ and $\bf r$ are defined by standard sequences. We will also
prove that the algebra $\Upsilon$ is not limited to atoms only:
${\bf 1}\setminus \bigcup_{i=1^\infty} {\bf a}_i)\not = {\bf 0}$.
By definition the element
${\bf e}={\bf 1}\setminus\bigcup_{i=1}^{\infty} {\bf a}_i$
is infinitely divisible, i.e. for any non-zero element
${\bf x}\subseteq {\bf e}$ a decomposition
${\bf x}={\bf x}_1\cup {\bf x}_2$ exists, where ${\bf x}_1\cap{\bf x}_2={\bf 0}$,
${\bf x}_1\not ={\bf 0}$ and ${\bf x}_2\not ={\bf 0}$.

Theorems~\ref{infini-div-1} and~\ref{nucl-ato-1} given below in Sections~\ref{nucl-2-pr-1}
and~\ref{atoms-count-1} will imply the main result
of this paper on decomposition of the maximal element of LV-algebra:

{\it The decomposition of the maximal element of $\Upsilon$ take place:
\begin{equation}
{\bf 1}=\cup_{i = 1}^{\infty}{\bf a}_i\cup{\bf e},
\label{main-decomposition-1}
\end{equation}
where $ {\bf a}_1, {\bf a}_2, \dots$ are all atoms of $\Upsilon$ and
${\bf e}$ is a non-zero infinitely divisible element.}

The decomposition (\ref{main-decomposition-1}) shows that any non-zero
LV-degree can be represented as a union of some atoms or it is an infinitely
divisible element, or it is a union of some atoms and non-zero infinitely
divisible element.

We show in Theorem~\ref{nucl-4} given below in Section~\ref{hyperimmune-2}
that the similar to (\ref{main-decomposition-1}) nontrivial decomposition
take place for LV-degree of all hyperimmune sequences:\footnote{i.e., for
indicator sequences of the hyperimmune sets of integer numbers.}
\begin{equation}
{\bf h}=\cup_{i = 1}^{\infty}{\bf h}_i\cup{\bf e},
\label{main-decomposition-2}
\end{equation}
where $ {\bf h}_1, {\bf h}_2, \dots$ are atoms and
${\bf e}$ is a non-zero infinitely divisible element generated by hyperimmune degrees.

\section{Network flows}\label{sec-net-1}

To construct the elements $\Upsilon$ generated by non-standard sequences,
we have to construct lower semicomputable semimeasures $P$ such that
$\bar P(\Omega\setminus (\bar R\cup C))>0$. 

We will construct some such semimeasure $P$ which will be represented as a flow 
over a certain network.

We will consider the set $\Xi$ of all finite binary sequences as a graph (tree)
whose vertices are sequences $x\in\Xi$ connected by edges of unit length
$(x,x0)$, $(x,x1)$. During the construction, we will add extra edges
$(x,y)$, where $x,y\in\Xi$, $x\subset y$, of length $l(y)-l(x)>1$.
For any edge $\sigma=(x,y)$ denote by $\sigma_1=x$ its starting vertex,
and by $\sigma_2=y$ its final vertex. A function $q(\sigma)$, which is defined
on all edges of unit length as well as on all extra edges, is called network if
\begin{equation} \label{net-1}
\sum\limits_{\sigma:\sigma_1=x} q(\sigma)\le 1
\end{equation}
for each $x\in\Xi$.
By $q$-flow we mean the minimal semimeasure $P$ such that
$P\ge R$, where the function $R$ (frame of the network flow) is defined as follows:
\begin{eqnarray}
R(\lambda)=1,
\label{net-base-1}
\\
R(y)=\sum\limits_{\sigma:\sigma_2=y}q(\sigma)R(\sigma_1)
\label{net-base-2}
\end{eqnarray}
for $y\not =\lambda$ (empty sequence).

It is easy to verify that the semimeasure $P$ can be defined as
\begin{eqnarray}
P(\lambda)=1,
\label{net-base-1a}
\\
P(y)=\sum\limits_{\sigma:\sigma_1\subset y\subseteq\sigma_2}q(\sigma)R(\sigma_1)
\label{net-base-2a}
\end{eqnarray}
for each $y$. The value $q(\sigma)$ can be interpreted as a portion of the
flow that goes from $x=\sigma_1$ to the vertex $y=\sigma_2$ along the edge $\sigma$.

We associate with any network $q$ the flow-delay function
$$
s(x)=1-q(x,x0)-q(x,x1).
$$
A network $q$ is called elementary if there exists an $n$ such that
$s(x)$ is defined for all $x$, $l(x)\le n$, the set $G^n$ of all extra edges is
finite and $l(\sigma_2)\le n$ for each $\sigma\in G^n$.
We assume that $q(x,x0)=q(x,x1)=\frac{1}{2}(1-s(x))$ for both edges of unit length
outgoing from $x$.

Any elementary network is a constructive object.
We will define a sequence of elementary networks gradually increasing $n$.

\subsection{Template~1}\label{nucl-2-pr}

We present the construction of a network $q$ depending on a recursive
predicate $B(i,\sigma)$, where $i$ is a positive integer number (task number),
$\sigma$ is an extra edge.

Let $p:{\cal N}\to{\cal N}$ be a total computable function such that for any
$i$, $p(n)=i$ for infinitely many $n$.\footnote{This function can be defined as follows.
Let $\langle i,j\rangle$ denotes
the order number of any pair $(i,j)$ of positive integer numbers under some
one-to-one corresponding between all positive integer numbers and all such pairs.
Define $p(\langle i,j\rangle)=i$ for all $(i,j)$.
}

Any extra edge $\sigma$ will refer to some task $i$ so that
$p(l(\sigma_1))=p(l(\sigma_2))=i$. We say that the edge $\sigma$ is of $i$th type.
The goal of the task $i$ will be to
draw extra edges $\sigma$ such that $B(i,\sigma)$ is satisfied and
that each infinite sequence $\omega$ passes
through one of these edges or the delay function would be equal to 1 on some
initial fragment of $\omega$.\footnote{An infinite sequence $\omega$ passes through
the edge $\sigma$ if $\sigma_2\subset\omega$.}

We associate with the predicate $B$ the function of setting an extra edge
\begin{equation}
\beta(x,n)=\min\{y:l(y)=n, p(l(y))=p(l(x)), B(p(l(x)),(x,y))\}.
\end{equation}
Here $\min$ is taken with respect to the natural linear orderings of all
finite binary sequences.
We suppose that $\min\emptyset$ is undefined. The pair $(x,\beta(x,n))$
will be drawn in $G$ as an extra edge.

Define a sequence of elementary networks by the mathematical induction on $n$.

Define $s(\lambda)=0$ and $G^0=\emptyset$.

Let $n\ge 1$ and $G^{n-1}$ and $q(\sigma)$ be defined for every
$\sigma\in G^{n-1}$, $s(x)$ be defined and for every $x$ such that $l(x)\le n-1$,
and $q(\sigma)=\frac{1}{2}(1-s(\sigma_1)$ for each $\sigma$ of unit length such that
$l(\sigma_2)=n$.

Let $G^{n-1}(i)$ be the set of all extra edges drawn by the task $i$ 
at steps $<n$. It should be $p(l(\sigma_1))=p(l(\sigma_2))=i$ for each 
$\sigma\in G^{n-1}(i)$.

We first introduce an auxiliary function $w(i,n)$.
The value of $w(i,n)$ is equal to the smallest $m$ such that $m\le n$, $p(m)=i$ and
$m>l(\sigma_2)$ for each extra edge $\sigma\in G^{n-1}(j)$ where $j<i$, i.e.,
for any extra edge $\sigma$ drawn during the processing of any task $j<i$.
Let us give the exact definition:
\begin{eqnarray}
w(i,n)=\min\{m: m\le n\&p(m)=i\&
\nonumber
\\
\forall j\forall\sigma
((j<i\&\sigma\in G^{n-1}(j)\rightarrow m>l(\sigma_2)).
\label{win-1}
\end{eqnarray}
We refer to $w(i,n)$ as to the initial step of a session for executing the task $i$.

The change in the value: $w(i,n)\not = w(i,n-1)$,
may occur due to the fact that at step $n$ some task $j<i$ draws its extra edge above
the level $w(i,n-1)$, and thus violates the condition
for the definition of $w(i,n-1)$.\footnote{By construction, if at a step $n$
of the induction some task draw a new extra edge $\sigma$ then $l(\sigma_2)=n$.
}
Lemma~\ref{gen-tech-1} will show that
this violation will occur at no more than a finite number of construction steps.

We will use a function $\rho(n)$ which is a parameter of the construction,
put $\rho(n)=(n+3)^2$.

The construction of step $n$ splits into three cases. Let $i=p(n)$.

{\it Case 1}. $w(p(n),n)=n$ (starting a new session for executing the task $i=p(n)$:
installing or reinstalling the task $i$).

In this case define $s(y)=1/\rho(n)$ for every $y$ such that $l(y)=n$ and
set $G^n=G^{n-1}$.

{\it Case 2.} $w(i,n)<n$ and $C_n(i)\not=\emptyset$, where
$C_n(i)$ is the set of all sequences $x$ that require processing, i.e., such that
$p(l(x))=i$, $w(i,n)\le l(x)<n$, $s(x)>0$,
$\beta(x,n)$ is defined and no extra edge in $G^{n-1}$ outgoes from $x$
(processing step of the task $i$).

In this case define
$G^n=G^{n-1}\cup\{(x,\beta(x,n)):x\in C_n(i)\}$
and $q((x,\beta(x,n)))=s(x)$ for each $x\in C_n(i)$.

If $s(x)<1$ then define $s(\beta(x,n))=0$ and
$
s(y)=s(x)/(1-s(x))
$
for all other $y$ of length $n$ such that
$x\subset y$ and $y\not=\beta(x,n)$.

If $s(x)=1$ then define $s(y)=0$ for every $y$ such that
$x\subset y$ and $l(y)=n$.

Define $s(y)=0$ for all other $y$ of length $n$.

{\it Case 3.} Cases 1 and 2 do not occur.
In this case define $s(x)=0$ for all $x$ of length $n$ and $G^n=G^{n-1}$.

After all, define $q(\sigma)=\frac{1}{2}(1-s(\sigma_1))$ for each $\sigma$ of
unit length such that $l(\sigma_1)=n$.

This concludes the description of the induction step.

Define $G=\cup_n G^n$ and $G(i)=\cup_n G^n(i)$ for any $i$. By the construction 
$s(x)$ is defined for each $x$ and $0\le s(x)\le 1$, 
$q(\sigma)$ is defined for each $\sigma\in G$
and $q(\sigma)=\frac{1}{2}(1-s(\sigma_1)$ for each $\sigma$ of unit length.

Lemmas~\ref{flow-delay-values-1}--\ref{two-edges-1} below present the simplest
properties of the construction.

\begin{lemma}\label{flow-delay-values-1}
The values of the flow delay function $s$ are $0$ or rational
numbers of type $\frac{1}{M}$, where $M$ is a positive integer number.
\end{lemma}
{\it Proof.}
By Case 1 at step $n$ we define $s(x)=\frac{1}{\rho(n)}$ for each $x$ such that
$l(x)=n$. By induction on $n$, in Case 2 if $s(x)=\frac{1}{M}$
for some $M>1$ then $s(y)=\frac{1}{M-1}$ for each $y$ such that $x\subset y$
and $y\not=\beta(x,n)$, also, $s(\beta(x,n))=0$. If $s(x)=1$ then $s(y)=0$
for each $y$ such that $x\subset y$ and $l(y)=n$.
$\Box$

The $q$-flow $P$ is lower semicomputable semimeasure by definition
(\ref{net-base-1a})--(\ref{net-base-2a}).

\begin{lemma}\label{two-edges-1}
There cannot be two overlapping extra edges $(x,y), (x',y')\in G$ such that
$x'\subset x\subset y'$ and $l(y')<l(y)$.
\end{lemma}
{\it Proof.} Assume that such a pair of overlapping extra edges exists.
Let $i=p(l(x))$ and $i'=p(l(x'))$. Evidently $i\not =i'$.
By the construction the extra edge $(x',y')$ was drawn at step $n'=l(y')$
and the extra edge $(x,y)$ was drawn at the later step step $n=l(y)$, where
$n>n'$ and $i<i'$.

There are two mutually exclusive cases. If $n''=l(x)=w(i,n'')$, i.e.,
the task $i$ was installed (or reinstalled) at the step $n''$. Then the pair $(x',y')$
cannot be an extra edge, since it should be added to $G$ at the later step $n'>n''$
by the task $i'>i$, that leads to the contradiction.

Assume that $n''=l(x)>w(i,n'')$, i.e., the task $i$ is processed at step $n''$.
Then Case 2 holds at step $n$ and $s(x)>0$. In this case, some extra edge
$\sigma$ such that $l(\sigma_2)=n''$ have to be drawn by the task $i$ at the step $n''$.
Then the contradiction is obtained, since the extra edge $(x',y')$ should be added to $G$
at the later step $n'>n''$ by the task $i'>i$.
The resulting contradiction proves the lemma. $\Box$

The next lemma shows that each task leads to the installation of new extra
edges only at a finite number of steps.

By construction $w(i,n+1)\ge w(i,n)$ for every $n$. 
Let $w(i)=\lim_{n\to\infty} w(i,n)$.
\begin{lemma}\label{gen-tech-1}
$G(i)$ is finite and $w(i)<\infty$ for each $i$.
\end{lemma}
{\it Proof.} Note that if $G(j)$ is finite for every $j<i$,
then $w(i)<\infty$. Therefore, it suffices to prove that $G(i)$ is finite for
each $i$. Assume the opposite. Let $i$ be the minimal for which
$G(i)$ is infinite. Since $G(j)$ is finite for each $j<i$, $w(i)<\infty$.

For any $x$ such that $l(x)\ge w(i)$, let $k$ be the maximal such that 
$\sigma_1=x^k$ and $l(\sigma_2)\le l(x)$ for some edge $\sigma\in G(i)$. 
This extra edge can be drawn by Case 2, where $s(x^k)>0$. By Lemma~\ref{flow-delay-values-1} 
$s(x^k)=1/M$, where $M$ is an integer number such that 
$M\ge 1$. If no such edge exists then set $k=w(i)$. Define
\[
  K(x)=
  \left\{
    \begin{array}{l}
     \rho(w(i)) \mbox{ if } l(x)\le w(i) \mbox{ or } k=w(i),\\
     M-1 \mbox{ if } l(x)>w(i)\mbox{ and } k>w(i), \mbox{ where }s(x^k)=1/M.\\
    \end{array}
  \right.
\]
Since $K(x)\ge K(y)$ for every $x$ and $y$ such that $x\subset y$, and,
moreover, if $K(x)>K(y)$ then $K(x)>K(z)$ for each
$z$ such that $x\subset z$ and $l(z)=l(y)$, the function
$$
\hat K(\omega)=\min\{n:\forall k\ge n (K(\omega^k)=K(\omega^n))\}
$$ 
is defined for each infinite $\omega\in\Omega$ and it is continuous.  
Since $\Omega$ is compact, it is upper bounded by some number $m$. Then
$K(x)=K(x^m)$ for every $x$ such that $l(x)\ge m$.

If at some step $n\ge m$ an extra edge $\sigma$ will be drawn by task $i$, where
$l(\sigma_1)>m$, then by Case 2 $K(y)<K(\sigma_1)$ for every $y$ of length $n$ 
such that $\sigma_1\subset y$. 
Therefore, the existence of such $m$ contradicts to the assumption of
infinity of $G(i)$. The lemma is proved. $\Box$

The support set of a semimeasure $P$ is defined as
$$
E_P=\{\omega\in\Omega:\forall n(P(\omega^n)>0)\}.
$$
It is easy to see that $\bar P(E_P)=\bar P(\Omega)$.

A sequence $\alpha\in\Omega$ is called $i$-extension of a finite sequence $x$
if $x\subset\alpha$ and $B(i,(x,\alpha^n))$ is satisfied for almost all $n$.

Note that if $\sigma\in G(i)$ is an extra
edge of the $i$th type then $B(i,\sigma)$ is satisfied.

\begin{lemma}\label{exten-1}
Let $\omega\in E_P$ and for any initial fragment $\omega^n$ of the sequence
$\omega$ there is an $i$-extension. Then $\omega$ passes through an extra edge of
the $i$th type (i.e., $\sigma_2\subset\omega$ for some $\sigma\in G(i)$).
\end{lemma}
{\it Proof.} By definition $P(\omega^n)\not =0$ for all $n$.
By Lemma~\ref{gen-tech-1}, there is a maximal $m$ such that $p(m)=i$ and $s(\omega^m)>0$.
Since $\omega^m$ has an $i$-extension and $s(\omega^m)>0$,
by Case 2 of the construction, an extra edge $(\omega^m,y)$ will be drawn on some
step $n$, where $l(y)=n$. Assume that $y\not\subset\omega$. If $s(\omega^m)<1$
then $s(\omega^n)>0$ that contradicts to the choice of $m$.
Let $s(\omega^m)=1$. Since $m\ge w(i,n)$ and by Lemma~\ref{two-edges-1},
no extra edge $\sigma$ exists such that $\sigma_1\subset\omega^m\subset\sigma_2$.
Then $P(\omega^{m+1})=0$ that is a contradiction with the
assumption of the lemma. Hence, $y\subset\omega$.
$\Box$

A semimeasure $P$ is continuous if $\lim\limits_{n\to\infty}P(\omega^n)=0$ for each
infinite sequence $\omega$. We give some sufficient condition for the continuity
of the $q$-network flow $P$.

A number $n$ separates the set $D$ of edges if $l(\sigma_1)\ge n$
or $l(\sigma_2)<n$ for each edge $\sigma\in D$.
\begin{lemma} \label{contin-1}
Let $q$ be a network. The $q$-flow is continuous if the set of extra edges
is separated by an infinite set of numbers, and $q(x,x0)=q(x,x1)$
for each $x\in\Xi$.
\end{lemma}
{\it Proof.} Let $P$ be the $q$-flow and a number $n$ separates the set
of extra edges. Then
$$
P(x)=R(x)=q(x^{n-1},x)R(x^{n-1})\le q(x^{n-1},x)P(x^{n-1})
$$
for each $x$ of length $n$. By~(\ref{net-1}) and by the assumption of the lemma
$q(x^{n-1},x)\le 1/2$ for all $x$ and $n$. Then
$P(\omega^n)\le(1/2)P(\omega^{n-1})$ for each $n$ separating the set of
extra edges. Since there are infinitely many of such $n$, we have
$\lim\limits_{n\to\infty} P(\omega^n)=0$, i.e., the semimeasure $P$
is continuous. $\Box$

The following corollary of Lemma~\ref{contin-1} takes place.
\begin{corollary}
Let $P$ be the flow through the network $q$ defined by Template 1.
Then the semimeasure $P$ is continuous.
\end{corollary}
{\it Proof.}
To apply Lemma~\ref{contin-1} to the semimeasure $P$, it suffices to note
that the number $w(i)$ separates $G$ for each $i$.
$\Box$

\begin{lemma}\label{nontriv-1a}
$\bar P({\bf 1})>0$.
\end{lemma}
{\it Proof.} Let us estimate $\bar P(\Omega)$ from below. Let $i=p(n)$. Define
$$
S_n=\sum\limits_{u:l(u)=n}R(u)-
\sum\limits_{\sigma:\sigma\in G,l(\sigma_2)=n}q(\sigma)R(\sigma_1).
$$
From the definition of the frame,
\begin{eqnarray}
\sum\limits_{u:l(u)=n+1}R(u)=\sum\limits_{u:l(u)=n}(1-s(u))R(u)+
\sum\limits_{\sigma:\sigma\in G,l(\sigma_2)=n+1}q(\sigma)R(\sigma_1).
\label{RR-2a}
\end{eqnarray}

Consider the case where $w(p(n),n)<n$.

If there is no edge $\sigma\in G$ such that $\sigma_1\in C_n(i)$ and $l(\sigma_2)=n$,
then $S_{n+1}\ge S_n$. Now, let $C_n(i)\not=\emptyset$. Define
$$
\Phi(\sigma,u)\Longleftrightarrow\sigma_1\in C_n(i)\&l(\sigma_2)=l(u)\&
\sigma_1\subseteq u\&u\not =\sigma_2.
$$
If $s(\sigma_1)=1$ for $\sigma_1\in C_n(i)$ then
$\sum\limits_{u:l(u)=n,\sigma_1\subseteq u}s(u)R(u)=0$.

By the construction, the value $s(x)$ defines a portion of the delayed flow in 
the vertex $x$.
The rest portion $1-s(x)$ of the flow goes equally to the vertices $x0$ and $x1$.
The portion $s(x)$ of the delayed flow can be later directed to some $y$ such that $l(y)=n$
and $x\subset y$ only by Case 2 along an extra edge $(x,y)$, where $y=\beta(x,n)$
outgoing from $x$ and drawn on the step $n$. We will show that at step $n$
the portion of the newly delayed flow in all $u$ of length $n$ such that
$x\subseteq u$ and $u\not=y$ does not exceed the portion of the previously delayed
flow at the vertex $x$ and directed along the edge $(x,y)$, this part of the flow
is no longer delayed by the task $i$ in the current session.

The following chain of equalities and inequalities take place:
\begin{eqnarray}
\sum\limits_{u:l(u)=n}s(u)R(u)=
\nonumber
\\
\sum\limits_{\sigma:\sigma_1\in C_n(i), s(\sigma_1)<1}
\sum\limits_{u:l(u)=n,\sigma_1\subseteq u}s(u)R(u)=
\nonumber
\\
\sum\limits_{\sigma:\sigma_1\in C_n(i),s(\sigma_1)<1}s(\sigma_2)
\sum\limits_{u:\Phi(\sigma,u)}R(u)=
\nonumber
\\
\sum\limits_{\sigma:\sigma_1\in C_n(i),s(\sigma_1)<1}
\frac{s(\sigma_1)}{1-s(\sigma_1)}
\sum\limits_{u:\Phi(\sigma,u)}R(u)\le
\nonumber
\\
\sum\limits_{\sigma:\sigma_1\in C_n(i),s(\sigma_1)<1}s(\sigma_1)R(\sigma_1)=
\nonumber
\\
\sum\limits_{\sigma:\sigma_1\in C_n(i),s(\sigma_1)<1}q(\sigma)R(\sigma_1)=
\nonumber
\\
\sum\limits_{\sigma:\sigma\in G,l(\sigma_2)=n}q(\sigma)R(\sigma_1).
\label{chain-1}
\end{eqnarray}
Here we have used the inequality
\begin{equation}\label{cont-flow-1}
\sum\limits_{u:\Phi(\sigma,u)}R(u)\le (1-s(\sigma_1))R(\sigma_1)
\end{equation}
for each $\sigma\in G^n$ such that $\sigma_1\in C_n(i)$ and $s(\sigma_1)<1$.
Inequality (\ref{cont-flow-1}) takes place, since the sum on the left is equal 
to the flow through the set of vertices $\{u:\Phi(\sigma, u)\}$, and the value 
from the right-hand side of the inequality is equal
to the value of the flow outgoing from the vertex $\sigma_1$,
except for its part passing through an extra
edge $\sigma$. By Lemma~\ref{two-edges-1} there cannot be an edge $\sigma'\in G$
overlapping with $\sigma$, i.e.,
such that $\sigma'_1\subset\sigma_1\subset\sigma'_2$ and $l(\sigma'_2)<l(\sigma_2)$.
Therefore, no extra portion of the flow from some vertex $\sigma'_1\subset\sigma_1$
cannot go through $\sigma_1$ and thus increase the flow to
$\{u:\Phi(\sigma, u)\}$.

Combining the resulting estimate with (\ref{RR-2a}),
we get $S_{n+1}\ge S_n$.

Consider now the case where $w(p(n),n)=n$. Then
$$
\sum\limits_{u:l(u)=n}s(u)R(u)\le 1/\rho(n)=1/(n+3)^2.
$$
Combining this inequality with 
(\ref{RR-2a}),
we get $S_{n+1}\ge S_n-1/(n + 3)^2 $. From here and from $S_0=1$ we get
$$
S_n\ge 1-\sum\limits_{i=1}^{\infty}1/(i+3)^2\ge\frac{1}{2}
$$
for all $n$. Since $P\ge R$,
$$
\bar P(\Omega)=\inf\limits_{n}\sum\limits_{l(u)=n} P(u)\ge
\inf\limits_n S_n\ge\frac{1}{2}.
$$
The lemma is proved. $\Box$

\section{Applications of Template 1}\label{appl-1}

In this section we present two applications of Template 1.

\subsection{Nonstochastic Turing degrees}\label{app-1}

We will prove that ${\bf 1}\setminus ({\bf r}\cup{\bf c})\not = {\bf 0}$.

Let $\{F_i\}$ be the uniformly computable sequence of all computable operators
We will assume that this sequence is modified by (\ref{modify-operator-1}) of
Section~\ref{prelimin-1} such that some output
$\tilde F_(x)\subseteq F_i(x)$ is obtained in $l(x)$ steps of computation and
the length of this output does not exceed the length $l(x)$ of the input $x$,
$\tilde F_i(\omega)=F_i(\omega)$ for each infinite $\omega$.

We also assume that for any computable operator $F$ there are infinitely many $i$
such that $F_i=F$.\footnote{To define such a sequence, redefine a sequence
of all computable operators $F_i$ as follows.
For any $i$, define $F'_{\langle i,j\rangle}=F_i$ for all $j$.
As before, the new sequence of operators will be denoted by $F_i$. Thus, for
any number $i$ of a computable operator $F_i$, one can enumerate an infinite
sequence of its other numbers.}

Define
\begin{eqnarray*}
B(i,\sigma)\Longleftrightarrow l(\tilde F_i(\sigma_2))>\sigma_1+i,
\end{eqnarray*}
where $\tilde F_i$ is the modified by (\ref{modify-operator-1}) computable
operator and the finite sequence $\sigma_1$ (the starting point of the edge $\sigma$)
is identified with its number in the natural numbering of the set $\Xi$.

\begin{theorem} \label{nontriv-1b}
For any infinite sequence $\omega$ from the support set of the semimeasure $P$ and
for any computable operator $F$, if $F(\omega)$ is infinite then the sequence
$F(\omega)$ is not Martin-L\"of random with respect to the uniform measure.
\end{theorem}
{\it Proof.} Note that if $F(\omega)$ is infinite and $F_i=F$, then for each
initial fragment of the sequence $\omega$ there is an $i$-continuation.
By Lemma~\ref{exten-1} for each such $i$, there is an edge $\sigma\in G(i)$
lying on $\omega$. For any $i$ define an open set
$$
U_i=\cup_{\sigma\in G(i)}\Gamma_{\tilde F_i(\sigma_2)}.
$$
Since $l(\tilde F_i(\sigma_2))>\sigma_1+i$ for $\sigma\in G(i)$,
$$
L(U_i)\le\sum\limits_{\sigma\in G(i)}2^{-\sigma_1-i}\le 2^{-i},
$$
where $L$ is the uniform measure. Define $U'_i=\cup_{j>i}U_i$, $L(U'_i)\le 2^{-i}$.
We have proved that $\{U'_i\}$ is Martin-L\"of test of randomness with respect
to the uniform measure. Since for any infinite $\omega$, $F(\omega)=\tilde F_i(\omega)$ 
for infinitely many $i$, $F(\omega)\in\cap_i U'_i$.
Thus, the sequence $F(\omega)$ is not Martin-L\"of random with respect to
the uniform measure $L$.
$\Box$

\begin{corollary}\label{nontriv-1b-cor-1}
$\bar P$-almost every infinite sequence $\omega$ cannot be Turing equvalent to
a sequence which is Martin-L\"of random with respect to some computable measure.
The a priory measure of all such sequences is positive.
\end{corollary}
{\it Proof.} The set of all computable sequences is countable. The continuity of
the semimeasure $P$ implies that $\bar P$-almost every sequence from its support set
is non-computable.
By~\cite{ZvL70} (Theorem~3.1), each non-computable
sequence Martin-L\"of random with respect to some computable measure is
algorithmically equivalent to a sequence which is Martin-L\"of random with respect
to the uniform measure. Therefore, $\bar P$-almost every sequence $\omega$ from
the support set of the semimeasure $P$ cannot be Turing equvalent to a sequence
Martin-L\"of random with respect to some computable measure.

Since the semimeasure $P$ is lower semicomputable, $cM\ge P$ for some constant $c$. Then
$\bar M(E_P)\ge\bar P(E_P)>0$.
$\Box$

\subsection{Infinitely divisible element}\label{nucl-2-pr-1}

We will construct a non-zero infinitely divisible element ${\bf e}\in\Upsilon$
which does not contain any atom. In order to do this, we apply Template 1
with specific recursive predicate $B(i,\sigma)$.

We will use the numbering of all pairs $\langle i,x\rangle$, where $i$ is a number and
$x$ is a finite sequence.\footnote{Recall that we identify finite sequence
and positive integer numbers.} The inverse functions also exist:
$[\langle i,x\rangle]_1=i$ is a task number, and the sequence
$[\langle i,x\rangle]_2=x$ is a candidate for processing.
Let $p(n)$ be such that for any $i$, $p(n)=i$ for infinitely many $n$.

We say that a sequence $z$ of length $n$ is $i$-discarded by an edge $\sigma\in G(i)$
at step $n$, where $i=[p(n)]_1$, if
$l(z)=l(\sigma_2)=n$ and $\tilde F_i(\sigma_2)\subseteq z$. Let $D_n(\sigma)$ be the set
of sequences of length $n$ which are $i$-discarded by the extra edge $\sigma$.

We slightly modify Case 2 of Template 1 to avoid collision between new extra edges
drawn at any step $n$ and the sequences discarded at step $n$. Now, at any step $n$, 
at most one extra edge $\sigma$ will be drawn in $G$, where $\sigma_1\in C_n(i)$.
Other elements of this set will be processed on later steps one by one. This edge $\sigma$
defines the set $D_n(\sigma)$ of $i$-discarded sequences of length $n$ such that
$D_n(\sigma)\cap\{z:l(z)=n, \sigma_1\subseteq z\}=\emptyset$.

We use the same sequence of all uniformly computable operators $\{\tilde F_i\}$
as in Section~\ref{app-1}.

Define the recursive predicate
\begin{eqnarray}
B(i,\sigma)\Longleftrightarrow \tilde F_i(\sigma_2)\not\subseteq\sigma_2\&
\sum\limits_{z:z\in D_n(\sigma)}R(z)\le 2^{-(\sigma_1+3)},
\label{rel-2}
\end{eqnarray}
where $R$ denotes the frame of the $q$-flow constructed in $n-1$ steps
and $\tilde F_i$ is modified by (\ref{modify-operator-1}).

{\it Modification of Case 2.}

{\it Case 2.} $w(i,n)<n$ and $C_n(i)\not=\emptyset$, where $i=[p(n)]_1$,
$C_n(i)$ is the set of all $x$ that requires processing, i.e., such that
$p(l(x))=i$, $w(i,n)\le l(x)<n$, $s(x)>0$, $\beta(x,n)$ is defined
and there is no extra edge in $G^{n-1}$ outgoing from $x$
(processing step of the task $i$).

If $x=[p(n)]_2\in C_n(i)$ then define
\begin{eqnarray*}
G^n=G^{n-1}\cup\{(x,\beta(x,n))\}.
\end{eqnarray*}
and $q((x,\beta(x,n)))=s(x)$. If $s(x)<1$ then define $s(\beta(x,n))=0$ and
$s(y)=s(x)/(1-s(x))$ for all other $y$ of length $n$ such that
$x\subset y$. If $s(x)=1$ then define $s(y)=0$ for these $y$.

For any sequence $z$ of length $n$, which is $i$-discarded by the extra edge
$\sigma=(x,\beta(x,n))$, define $s(z)=1$.

Define $s(x)=0$ for all other $x$ of length $n$ and define
$q(\sigma)=\frac{1}{2}(1-s(\sigma_1))$
for all edges $\sigma$ of unit length such that $l(\sigma_1)=n$.

This modification does not change the basic properties of the Template 1.

Let $s$ be the flow delay function for the network $q$ and $P$
denotes the $q$-flow. The frame $R$ is defined using equalities
(\ref{net-base-1})--(\ref{net-base-2}).

Similarly to how it was done in Section~\ref{nucl-1-pr},
we can prove that $P$ is continuous lower semicomputable semimeasure.

We modify the proof of Lemma~\ref{nontriv-1a} for the case
where some sequences are discarded.
\begin{lemma} \label{nontriv-1b-2}
$\bar P({\bf 1})>0$.
\end{lemma}
{\it Proof.} Let us estimate from below the value of $\bar P(\Omega)$.
Consider
$$
S_n=\sum\limits_{u:l(u)=n}R(u)-
\sum\limits_{\sigma:\sigma\in G,l(\sigma_2)=n}q(\sigma)R(\sigma_1).
$$
From the definition of the frame, we have
\begin{eqnarray}
\sum\limits_{u:l(u)=n+1}R(u)=\sum\limits_{u:l(u)=n}(1-s(u))R(u)+
\label{RR-1b}
\\
\sum\limits_{\sigma:\sigma\in G,l(\sigma_2)=n+1}q(\sigma)R(\sigma_1).
\label{RR-2b}
\end{eqnarray}

In case $w(p(n),n)=n$
$$
\sum\limits_{u:l(u)=n}s(u)R(u)\le\rho(n)=1/(n+3)^2.
$$
Combining this inequality with (\ref{RR-1b})--(\ref{RR-2b}), where
the sum (\ref{RR-2b}) is equal to 0, we obtain
$S_{n+1}\ge S_n-1/(n+3)^2$.

Let $w(p(n),n)<n$ and $\sigma\in G$ such that $\sigma_1=[p(n)]_2$. Then
\begin{eqnarray}
\sum\limits_{u:l(u)=n}s(u)R(u)=
\nonumber
\\
\sum\limits_{u:l(u)=n,u\not\in D_n(\sigma)}s(u)R(u)+
\sum\limits_{u:l(u)=n,u\in D_n(\sigma)}R(u).
\label{represent-1}
\end{eqnarray}
The first sum in (\ref{represent-1}) is bounded similarly as (\ref{chain-1})
of the proof of Lemma~\ref{nontriv-1a}:
\begin{eqnarray}
\sum\limits_{u:l(u)=n,u\not\in D_n}s(u)R(u)\le
q(\sigma)R(\sigma_1).
\label{chain-1a}
\end{eqnarray}
The second sum is bounded as
\begin{eqnarray}
\sum\limits_{u:l(u)=n,u\in D_n(\sigma)}R(u)\le 2^{-(\sigma_1+3)}.
\label{rel-2a}
\end{eqnarray}
The bounds (\ref{chain-1a}) and (\ref{rel-2a})
implies the lower bound
$$
S_n\ge 1-\sum\limits_{i=1}^{\infty}1/(i+3)^2-
\sum\limits_{\sigma}2^{-(\sigma_2+3)}\ge\frac{1}{2}
$$
for all $n$. Since $P\ge R$, we have
$$
\bar P(\Omega)=\inf\limits_{n}\sum\limits_{l(u)=n} P(u)\ge
\inf\limits_n S_n\ge\frac{1}{2}.
$$
Lemma is proved. $\Box$

\begin{lemma}\label{not-rduc-1}
Any two different infinite sequences $\omega$ and $\alpha$ from
the set $E_P$ are not Turing reducible to each other.
\end{lemma}
{\it Proof.} Assume that $\alpha=F_i(\omega)$ for some $i$.
Since $\omega\not=\alpha$, $\tilde F_i(\omega^n)\not\subseteq\omega^n$ for all sufficiently
large $n$. Then by Lemma~\ref{exten-1}, for all sufficiently large $n>w(i)$
such that $[p(n)]_1=i$ an edge $\sigma$ exists such that $l(\sigma_2)=n$,
$\sigma_2\subset\omega$ and $B(i,\sigma)$ is satisfied, in particular,
$\sigma_1\in C_n(i)$. By construction $\sigma=[p(n)]_2$ for some of these $n$.
Then the sequence $\alpha^n$ will be $i$-discarded for some $n>w(i)$ and $s(\alpha^n)=1$
will be defined. No extra edge $\sigma'$ such that
$\sigma'_1\subset\alpha^n\subset\sigma'_2$ can be drawn at any step
$n'=l(\sigma'_2)>n$, and therefore, no extra portion of the flow
can go through the vertex $\alpha^n$. Indeed, any task $j<i$ cannot draw
extra edges on steps $n'>n$ since $w(i,n)=w(i)$. At the step $n$ the sessions of all tasks
$j>i$ are terminated, and on the later steps $n'>n$ the tasks $j>i$ can draw only extra
edges $\sigma'$ such that $l(\sigma'_1)>n$.

Hence, $q((\alpha^n,\alpha^{n+1}))=0$, and so, $\alpha\not\in E_P$.
This contradiction proves that $\alpha$ is not Turing reducible to $\omega$.
$\Box$

\begin{theorem}\label{infini-div-1}
There exists a non-zero infinitely divisible element ${\bf e}$ such that
$$
{\bf e}\cap({\bf r}\cup{\bf c})={\bf 0}.
$$
\end{theorem}
{\it Proof.}
Define ${\bf f}=[\bar E_P]$. Let ${\bf e}=i_P({\bf f})=[E]$.
Assume that ${\bf x}\subseteq{\bf e}$ and ${\bf x}\not={\bf 0}$. Take an
$X\in{\bf x}$ and put $X'=X\cap E$. Clearly, $\bar P(X')>0$.
Since $X'\subseteq E$, by Lemma~\ref{Rad-Nic} $\bar M(X')>0$.
Since $X'\subseteq E_P$, by Lemma~\ref{not-rduc-1}
any two sequences from $X'$ do not reducible to each other.
Let us represent $X'=X_1\cup X_2$, where $X_1\cap X_2=\emptyset$, $\bar P(X_1)>0$,
and $\bar P(X_2)>0$. Let ${\bf x}_1=[\bar X_1]$ and ${\bf x}_2=[\bar X_2]$.
Then ${\bf x}={\bf x}_1\cup {\bf x}_2$, ${\bf x}_1\not={\bf 0}$,
${\bf x}_2\not={\bf 0}$ and ${\bf x}_1\cap {\bf x}_2={\bf 0}$.
Hence, $\bf x$ cannot be an atom. Theorem is proved.
$\Box$

\subsection{Template~2}

In this section we present Template~2 which is a modification of Template~1
and which will be used to construct atoms of the algebra $\Upsilon$.
The modification given below does not affect the above properties of Template~1.

Let $p(n)$ and $\tilde p(n)$ be computable functions such that
for each pair of positive integer numbers $(i,k)$,
$p(n)=i$ and $\tilde p(n)=k$ for infinitely many $n$.

Any extra edge $\sigma$ corresponds to a task $i$, where
$p(l(\sigma_1))=p(l(\sigma_2))=i$. It also corresponds to some
subtask $(i,k)$, where $\tilde p(l(\sigma_1))=\tilde p(l(\sigma_2))=k$.

By induction on $n$, define a sequence of elementary networks.
and the sets of extra edges $G^n$.

Define $s(\lambda)=0$ and $G^0=\emptyset$. The induction hypothesis
is the same as for step $n$ of Template 1.


Consider an auxiliary function $w(i,n)$. The value of $w(i,n)$ is equal to
the least $m$ such that $m\le n$, $p(m)=i$ and $m>l(\sigma_2)$ for each extra edge
$\sigma$, which was drawn by a task $j<i$. In particular, for each $n'$
such that $w(i,n)\le n'\le n$ no task $j<i$ was processed. The formal
definition was given by (\ref{win-1}).
We refer to $w(i,n)$ as to the initial step of a session to process the task $i$.

The equality $w(i,n)=w(i,n-1)$ is violated (i.e. $w(i,n)\not =w(i,n-1)$)
only if some task $j<i$ has established an extra edge located above the
level $w(i,n-1)$, and thus it violates the condition (\ref{win-1})
for the definition of $w(i,n)$. Lemma~\ref{gen-tech-1} states that this can
only happen at a finite number of construction steps.

Define a family of equivalence relations between finite sequences depending
on the parameter $w$:
$$
x\sim_w y\Longleftrightarrow l(x)=l(y)\&\forall(w\le i\le l(x)\Longrightarrow x_i=y_i);
$$
Note that if $x\sim_w y$ then $x\sim_{w'} y$ for each $w'\ge w$.

For any edges $\sigma$ and $\sigma'$, define $\sigma\sim_w\sigma'$
if and only if $\sigma_1\sim_w\sigma'_1$ and $\sigma_2\sim_w\sigma'_2$.

Sometimes, we write $x\sim z$ instead of $x\sim_w z$, where $w=w(p(l(x)),l(x))$.


During the construction process, we will execute the task $i$ by executing
the subtasks $(i,k)$ in the order of their priority. If the edge is drawn
by subtask $(i,k)$ then we say that it also is drawn by the task $i$.

At any step $n$, let $z_{i,1,n},\dots, z_{i,2^{w(i,n)},n}$ be all finite binary sequences
$z$ of length $w(i,n)$ written out in the lexicographic order. We refer
to the sets $T_{z_{i,t,n}}=\{y:z_{i,t,n}\subseteq y\}$ as to subtrees of the tree $\Xi$
with the roots $z_{i,1,n},\dots, z_{i,2^{w(i,n)},n}$.

Given a set of extra edges $G$ let
$$
G(i)=\{\sigma:\sigma\in G\&p(l(\sigma_1))=p(l(\sigma_2))=i\}
$$
be the set of all extra edges drawn by the task $i$ and
$$
G(i,k)=\{\sigma\in G(i):\tilde p(l(\sigma_1))=\tilde p(l(\sigma_2))=k\}
$$
be the subset of all extra edges drawn by the subtask $(i,k)$.

By definition, the $w(i,k,n) $ is equal to the smallest $m$
such that $m\le n$ and the following conditions are satisfied. First, $p(m)=i$ and
$m>l(\sigma_2)$ for each extra edge $\sigma$ which was drawn by
the task $j<i$. Second, $m>l(\sigma_2)$ for each extra edge $\sigma$ that was
drawn by some subtask $(i,t)$, where $t<k$.
This means that at all steps $n'$ such that $w(i,k,n)\le n'\le n$
any subtask $(j,t)$, where $j<i$ or $j=i$ and $t<k$, did not draw new extra edges.

Let us give the exact definition. For $k\le 2^{w(i,n)}$ define
\begin{eqnarray}
w(i,k,n)=\min\{m: m\le n\&p(m)=i\&\tilde p(m)=k\&
\nonumber
\\
\forall j\forall\sigma ((j<i\&\sigma\in G^{n-1}(j)\rightarrow m>l(\sigma_2))\&
\\
\nonumber
\forall t\forall\sigma (1\le t<k\&\sigma\in G^{n-1}(i,t)\rightarrow m>l(\sigma_2))\}.
\label{wikn-2}
\end{eqnarray}
Assume that $\min\emptyset=\infty$.

We say that $w(i,k,n)$ is the initial step of the sub-session for executing
the subtask $(i,k)$. By definition, $w(i,k,n)\ge w(i,n)$ for each $i$ and $k$.
Any session for the execution of any task $i$ consists of sub-sessions $(i,k)$,
which are executed in order of their priority.

Violation of the equality $w(i,k,n)=w(i,k,n-1)$ can occur because
$w(i,n)\not =w(i,n-1)$ or for $w(i,n)=w(i,n-1)$ because some subtask
$(i,t)$, where $t<k$, draws an extra edge above the level
$w(i,k,n-1)$ and thus violates the condition for the definition of
$w(i,k,n-1)$. It will be shown below that such a violation will occur at
no more than a finite number of steps $n$ such that $w(i,n)=w(i,n-1)$.

We define a network, depending on a recursive predicate
$B(i,\sigma)$, where $i$ is a positive integer number (task number),
$\sigma$ is an extra edge.

The goal of the task $i$ is the same as for Template 1 -- to draw extra edges
$\sigma$ such that each infinite sequence $\omega$ from the support set
of the corresponding network flow passes through one of these edges.
Each such edge $\sigma$ should satisfy the predicate $B(i,\sigma)$.

There is an additional requirement. In order for the corresponding flow to define
an atom of the algebra $\Upsilon $, in the modified construction the value of
the flow through any two edges $\sigma$ and $\sigma '$ such that $\sigma'\sim\sigma$
should be the same. Therefore, when the edge $\sigma$ is drawn by the task $i$,
all the edges $\sigma'\sim\sigma$ and $\sigma'\not=\sigma$ become dependent on it.
All assignments on these edges should to mimic the assignments on $\sigma$.
In this case, a collision may occur if we try to simultaneously make assignments
of the task $i$ by Case 2 for another edge, which is located in a different subtree.
In order to avoid a collision when setting the edges of the task $i$, at step $n$
we split the process of executing the task $i$ into subtasks $(i,k)$, where
$k=1,\dots 2^{w(i,n)}$.

At any step $n$ of the construction we execute the subtask $(i,k)$ into the subtree
$T_{z_{i,k,n}}$ (we call it the leading subtree), where $k=\tilde p(n)$,
and duplicate all actions in all other subtrees $T_{z_{i,t,n}}$ for $t\not=k$
(we call them dependent subtrees). We will perform the task $i$ for the other
subtrees in subsequent steps in order of their priority, still repeating all
the assignments in the remaining subtrees.

By the construction below if the subtask $(i,k)$ is executed at step $n$,
then for each extra edge $\sigma$ which was drawn by any subtask $(i,t)$ where $t<k$
it will be $l(\sigma_2)<w(i,k,n)$ and, therefore, the equality
$w(i,t,n)=w(i,t,n-1)$ will not be violated for $t<k$.
If a new edge will be drawn in this subtree then
all subtasks $(i,t)$ corresponding to subtrees with lower priority $t>k$ will
be terminated and the equalities $w(i,t,n)=w(i,t,n-1)$ will be violated.
These subtasks should be reinstalled on later steps.

Each extra edge $\sigma$ will refer to a task $i$ such that
$p(l(\sigma_1))=p(l(\sigma_2))=i$ and to the subtask $(i,k)$,
where $\tilde p(l(\sigma_1))=\tilde p(l(\sigma_2))=k$.

The predicate $B$ defines the function of setting an extra edge
$$
\beta(x,n)=\min\{y:l(y)=n, p(l(y))=p(l(x)),B(p(l(x)),(x,y))\}.
$$
We assume that $\min\emptyset$ is undefined.

Define $s(\lambda)=0$ and $G^0=\emptyset$.

At any step $n$, let $i=p(n)$ and $k=\tilde p(n)$.

Let $n\ge 1$ and $G^{n-1}$ and $q(\sigma)$ be defined for every
$\sigma\in G^{n-1}$, $s(x)$ be defined for every $x$ such that $l(x)\le n-1$,
and $s(x)=s(x')$ for every $x$ and $x'$ of length $<n$ such that $x\sim_{w(i,n)}x'$.
Let also, $q(\sigma)=\frac{1}{2}(1-s(\sigma_1)$ for each $\sigma$ of unit length such that
$l(\sigma_2)=n$.

If $k>2^{w(i,n}$ then define $s(x)=0$ for each $x$ of length $n$ and $G^n=G^{n-1}$
and go to the next step. If $k\le 2^{w(i,n)}$ then go to the definitions below.
 
The construction of any step $n$ splits into three cases:

{\it Case 1}. $w(i,k,n)=n$ (starting a new sub-session for executing the subtask
$(i,k)$, where $i=p(n)$ and $k=\tilde p(n)$: first installing or reinstalling
of the subtask $(i,k)$).

Define $s(y)=1/\rho(n)$ for all $y$ such that $l(y)=n$ and set $G^n=G^{n-1}$,
where $\rho(n)=(n+3)^2$.

{\it Case 2.} $w(i,k,n)<n$ and $C_n(i,k)\not=\emptyset$, where
$C_n(i,k)$ denotes the set of all sequences $x$
which should be processed by subtask $(i,k)$, i.e.,
such that $z_{i,k,n}\subseteq x$, $p(l(x))=i$ and $\tilde p(l(x))=k$, 
$w(i,k,n)\le l(x)<n$, $s(x)>0$,
$\beta(x,n))$ is defined and there is no extra edge from $G^{n-1}$ outgoing from $x$
(the subtask $(i,k)$ processing step).

In this case, we make identical assignments in all subtrees $T_{z_{i,t,n}}$,
$1\le t\le 2^{w(i,n)}$, which repeat the assignments in the leading subtree
$T_{z_{i,k,n}}$ of the subtask $(i,k)$:

1) For any $x\in C_n(i,k)$ define $s(z)=0$ for each $z$ such that
$z\sim_{w(i,n)}\beta(x,n)$ and $q(\sigma)=s(x)$ for every $\sigma$ such that
$\sigma\sim_{w(i,n)}(x,\beta(x,n))$.
We add all these edges $\sigma$ to $G^{n-1}$ and get $G^n$.

2) If $s(x)<1$ then for any $y$ such that $x\subset y$, $l(y)=n$ and
$y\not=\beta(x,n)$ define  $s(z)=s(x)/(1-s(x))$ for each $z$ such that
$z\sim_{w(i,n)} y$.

3) If $s(x)=1$ then define $s(z)=0$ for all these $z$.

4) Define $s(z)=0$ for all other $z$ of length $n$.

{\it Case 3.} Cases 1 and 2 do not occure. In this case define $s(x)=0$ for each
$x$ of length $n$, and define $G^n=G^{n-1}$.

After all, define
$
q(\sigma)=\frac{1}{2}(1-s(\sigma_1))
$
for each $\sigma$ of unit length such that $l(\sigma_1)=n$.

This concludes the description of the induction step.

Define $G=\cup_n G^n$ and $G(i)=\cup_n G^n(i)$, $G(i,k)=\cup_n G^n(i,k)$ 
for any $i$ and $k$. 
Let $w(i)=\lim_{n\to\infty}w(i,n)$ and $w(i,k)=\lim_{n\to\infty}w(i,k,n)$.

The analogs of the Lemmas~\ref{flow-delay-values-1}--\ref{gen-tech-1}
also take place for the modified construction.
In particular, any task $i$ is processed only at a finite number
of steps and the set $G(i)$ is finite and $w(i)<\infty$ for all $i$.

The next lemma states that any subtask $(i,k)$ is processed only at
a finite number of steps.
\begin{lemma} \label{gen-tech-1a}
The set $G(i,k)$ is finite and $w(i,k)<\infty$ for all $i$ and $k\le 2^{w(i)}$.
\end{lemma}
{\it Proof.} By Lemma~\ref{gen-tech-1} $w(i)<\infty$. Then
$w(i,n)=w(i)$ for all $n\ge n'$ for some $n'$.
Further, following the proof of Lemma~\ref{gen-tech-1}, where the function $w(i,n)$
is replaced with $w(i,k,n)$, we show that
$w(i,k,n)\not =w(i,k,n-1)$ only for a finite number of different $n\ge n'$.
$\Box$

The following duplication property takes place.
\begin{lemma}\label{dup-1}
For any $i$, $q(\sigma)=q(\delta)$ for each $\sigma,\delta\in G$
such that $\sigma\sim_{w(i)}\delta$. 
\end{lemma}
{\it Proof.} Since $w(i)=\lim_{n\to\infty}w(i,n)$,
only tasks $j\ge i$ can draw the extra edges $\sigma\in G$ on steps $n\ge w(i)$,
where $l(\sigma_1)\ge w(i)$.
Assume that the extra edges $\sigma,\delta$ be drawn by some subtask $(j,k)$,
where $j\ge i$.
By definition a single extra edge $\sigma'\in G$ exists in the leading subtree
such that $z_{j,k,n}\subset\sigma'$, $\sigma\sim_{w(j,n)}\sigma'$
and $q(\sigma)=q(\sigma')$, where $n=l(\sigma_2$ and $j=p(n)$ and $k=\tilde p(n)$.
Similarly, a single extra edge $\delta'\in G$ exists such that
$\delta\sim_{w(j,n)}\delta'$ and $q(\delta)=q(\delta')$.

Since $\sigma\sim_{w(i)}\delta$ and $w(j,n)\ge w(i)$, we have
$\sigma\sim_{w(j,n)}\delta$. Then $\sigma'=\delta'$ and
by the construction $q(\sigma)=q(\delta)$.
$\Box$

Let $R$ be the frame of the $q$-flow $P$. Clearly, the semimeasure $P$
is lower semicomputable.

By Lemma~\ref{contin-1} the semimeasure $P$ is continuous, since
for any $i$ the number $w(i)$ separates $G$.

The support set of a semimeasure $P$ is equal to
$$
E_{P}=\{\omega\in\Omega:\forall n(P(\omega^n)\not=0)\}.
$$

The following lemma will enable us to apply Kolmogorov 0 or 1 law to the measure
$\bar P$, where $P$ is the $q$-flow. Although this measure is not normalized, 
this does not lead to a loss
of generality, since the subsequent statements do not depend on the multiplicative factor.

For any $i$, let $f_i(\omega)=\omega_i$ be a random variable in the probability space
$(\Omega,\bar P)$.
\begin{lemma} \label{zero-one-law}
For any $n>w(i)$, the random variable $f_n$ does not depend on the random
variables $\{f_j:j\le w(i)\}$.
\end{lemma}
{\it Proof.} Since $w(i)=\lim_{n\to\infty}w(i,n)$,
only tasks $i'\ge i$ can draw the extra edges $\sigma\in G$ on steps $n\ge w(i)$,
where $l(\sigma_1)\ge w(i)$. From this it follows that
for $l(v)>w(i)$, the formula (\ref{net-base-2}) can be rewritten as
\begin{eqnarray}
R(v)=\sum\limits_{l(\sigma_1)\ge w(i),\sigma_2=v}q(\sigma)R(\sigma_1).
\label{short-1}
\end{eqnarray}
Assume that $v\sim_{w(i)} v'$ and $l(v)>w(i)$. These sequences belong to
some subtrees. 
By Lemma~\ref{dup-1} for any $\sigma\in G$ such that
$l(\sigma_1)\ge w(i)$ and $\sigma_2=v$ there exists $\sigma'\in G$
which belongs to the same subtree as $v'$ and
such that $\sigma'\sim_{w(i)}\sigma$ and $q(\sigma')=q(\sigma)$.
Clearly, $\sigma'_2=v'$. Using induction on recursive definition (\ref{short-1}),
we obtain
\begin{equation}\label{rela-1}
R(v)/R(v^{w(i)})=R(v')/R(v'^{w(i)}).
\end{equation}
Since there are no extra edges $\sigma$ such that $\sigma_1\subset z^{w(i)}\subset\sigma_2$,
the equality $P(z^{w(i)})=R(z^{w(i)})$ takes place.
Now, using the representation (\ref{net-base-1a})--(\ref{net-base-2a}), we will
prove the similar equality for $P$.

Let $l(y)>w(i)$ and $y\sim_{w(i)} z$. Then $y\sim_n z$ for each $n\ge w(i)$.
By Lemma~\ref{dup-1} for each extra edge $\sigma\in G$ such that
$y\subseteq\sigma_1\subset y\subseteq\sigma_2$,
an extra edge $\sigma'\sim_{w(i)}\sigma$ exists such that
$z\subseteq\sigma'_1\subseteq z\subseteq\sigma'_2$ and $q(\sigma)=q(\sigma')$.
From these and by (\ref{rela-1}), we obtain
\begin{equation}\label{relation-2}
P(y)/P(y^{w(i)})=P(z)/P(z^{w(i)})
\end{equation}
for all $y$ and $z$ such that $y\sim_{w(i)}z$. From this we obtain
$$
\bar P(y)/\bar P(y^{w(i)})=\bar P(z)/\bar P(z^{w(i)})
$$
for all $y$ end $z$ such that $y\sim_{w(i)}z$.

Therefore, the conditional probability
$$
\bar P(y|x)=\frac{\bar P(xy)}{\bar P(x)}
$$
does not depend on the choice of the initial fragment $x$ of the sequence $y$
for $l(x)=w(i)$ and $l(y)>w(i)$:

In particular, the random variables $f_j(\omega)=\omega_j$ do not depend on
the random variables $f_s(\omega)=\omega_s$ for $s\le w(i)<j$.
$\Box$

We define an atom consisting of nonstochastic Turing degrees.

Let $q$ be the network defined using the Template 2,
$G$ be the set of all extra edges, and $s$ be the corresponding flow delay function.

The following lemma is a corollary of Lemma~\ref{zero-one-law}.
\begin{lemma} \label{atom-el-1}
For any $A\subseteq\Omega$ containing, together with
each sequence, all sequences differing from it by a finite number of bits,
$\bar P(A)=0$ or $\bar P(A)=\bar P(\Omega)$.
\end{lemma}
{\it Proof.}\footnote{A specific for Cantor space $(\Omega,L)$ proof of this lemma 
see in Downey and Hirschfeldt~\cite{DH2010}, Theorem 1.2.4. See also,
Holzl and Porter~\cite{HoP2021}, Theorem 4.9.}
To apply the Kolmogorov 0 or 1 law, consider the independent random
variables $\tilde f_1,\tilde f_2,\dots$, where
$$
\tilde f_i(\omega)=f_{w(i)+1}(\omega)\dots f_{w(i+1)}(\omega)=
\omega_{w(i)+1}\dots\omega_{w(i+1)}
$$
and $f_i(\omega)=\omega_i$. Clearly, the random variables
$\tilde f_1(\omega),\tilde f_2(\omega),\dots$ generate the same $\sigma$-algebra
as the random variables $f_1(\omega),f_2(\omega),\dots$.

The set $A$ satisfying the condition of the lemma lies in the
$\sigma$-algebra generated by the set of independent random variables
$\tilde f_{k},\tilde f_{k+1},\dots$ for each $k$,
and therefore, it lies in the residual $\sigma$-algebra of the entire sequence
$\tilde f_1,\tilde f_2,\dots$. By Kolmogorov 0 or 1 law
$\bar P(A)=0$ or $\bar P(A)=\bar P(\Omega)$.
$\Box$

\begin{corollary}\label{atom-el-2}
There exists an atom of $\Upsilon$.
\end{corollary}
{\it Proof.}
Let ${\bf p}=[\bar E_P]$. By Lemma~\ref{nontriv-1a} $P({\bf p})>0$,
then ${\bf p}\not ={\bf 0}$. Define
${\bf d}=i_P({\bf p})$. By definition $\bar P(i_P({\bf p}))=\bar P({\bf d})$.
Then ${\bf d}\not ={\bf 0}$.

Assume that ${\bf d}={\bf a}\cup{\bf b}$, where ${\bf a}\not ={\bf 0}$,
${\bf b}\not ={\bf 0}$ and ${\bf a}\cap\bf b={\bf 0}$.
By Corollary~\ref{Rad-Nic} $\bar P({\bf a})>0$ and $\bar P({\bf b})>0$,
that contradicts Lemma~\ref{atom-el-1}.
This contradiction proves that ${\bf d}$ is an atom.
$\Box$

\subsection{Atom of nonstochastic Turing degrees}\label{nucl-1-pr}
 
Corollary~\ref{atom-el-2} shows that the network flow defined by Template 2
generate an atom regardless what predicate $B(i,\sigma)$ is used.
Specifying this predicate, we obtain the following theorem.

\begin{theorem}\label{single-atom-1}
There exists an atom ${\bf d}$ such that ${\bf d}\cap({\bf c}\cup{\bf r})={\bf 0}$.
\end{theorem}
{\it Proof.}
Let us specify the predicate:
\begin{eqnarray*}
B(i,\sigma)\Longleftrightarrow l(\tilde F_i(\sigma_2))>\sigma_1+i,
\end{eqnarray*}
where the finite sequence $\sigma_1$ (the starting point of the edge $\sigma$)
is identified with its order number in the natural numbering of the set $\Xi$.

The following statements are similar to Theorem~\ref{nontriv-1b}
and Corollary~\ref{nontriv-1b-cor-1} and their proofs are the same:

1) For any infinite sequence $\omega$ from the support set of the semimeasure $P$ and
for any computable operator $F$, if $F(\omega)$ is infinite, then the sequence
$F(\omega)$ is not Martin-L\"of random with respect to the uniform measure.

2) $\bar P$-almost every infinite sequence $\omega$ is not Martin-L\"of random
with respect to any computable measure.

From these statements ${\bf d}\cap({\bf c}\cup{\bf r})={\bf 0}$ follows.
$\Box$

\subsection{Decomposition into countable sequence of atoms}\label{atoms-count-1}

We will construct an infinite sequence of lower semicomputable semimeasures
$P_1, P_2,\dots$ which will define a sequence of pairwise different atoms
$\bf d_1, \bf d_2,\dots$.

We use the same sequence of all computable operators $\{F_i\}$ and their
modified versions $\{\tilde F_i\}$ as in Section~\ref{app-1}.

Let $\langle x_1,x_2,x_3\rangle$ be the number of a triple of natural numbers,
for some fixed computable one-to-one correspondence between all triples
$\langle x_1,x_2,x_3\rangle$ such that $x_1\not=x_2$, and all positive integer numbers.

The inverse functions $[\langle x_1,x_2,x_3\rangle]_t=x_t$, $t=1,2,3$, are also given.
The order number $i=\langle x_1,x_2,x_3\rangle$ of each such triplet will be a code
of some task $i$, where $x_1$ is the number of the computable operator, $x_2$ is
called task base, $x_3$ is task target.
By the main property of triplets numbering, any number $m$ cannot be both
the target and the base of the same task.

Let us define the networks $q_m$ for $m=1,2,\dots$.
We will execute the tasks what are common to all networks $q_m$.
At any step $n$ the task $i=p(n)$ will execute Template~2 for the network $q_{[i]_1}$,
which is the base of the task $i$ and discard some vertices
of the network $q_{[i]_2}$, which is the target of the task $i$.
All other networks remains unchanged at the step $n$.
One and the same network can serve as a base for some task at some steps and
as a target at other steps, but not at the same time.

For any $m$, let $G^n_m(i)$ be the set of all extra edges drawn by a task $i$ 
for a network $q_m$ at steps $\le n$, $G^n_m=\cup_i G^n_m(i)$. 
Since at any step of the construction only a finite number of extra edges 
can be drawn, for any $n$, $G^n_m=\emptyset$ for almost all $m$.

Definition (\ref{win-1}) of the function $w(i,n)$ is changed to
\begin{eqnarray*}
w(i,n)=\min\{n': n'\le n\&p(n')=i\&
\nonumber
\\
\forall m\forall j\forall\sigma
((j<i\&\sigma\in G_m^{n-1}(j)\rightarrow n'>l(\sigma_2)).
\end{eqnarray*}
and the definition (\ref{wikn-2}) of the function $w(i,k,n)$ is changed to
\begin{eqnarray*}
w(i,k,n)=\min\{n': n'\le n\&p(n')=i\&\tilde p(n')=k\&
\nonumber
\\
\forall m\forall j\forall\sigma ((j<i\&\sigma\in G_m^{n-1}(j)
\rightarrow n'>l(\sigma_2))\&
\\
\nonumber
\forall m\forall t\forall\sigma (1\le t<k\&\sigma\in G_m^{n-1}(i,t)
\rightarrow n'>l(\sigma_2))\}.
\end{eqnarray*}

We add a new element to construction of Template~2.
We say that a finite sequence $x$ of length $n$ is $i$-discarded by a sequence
$y$ (or by an edge $\sigma\in G^n_{[i]_1}$ such that $\sigma_2=y$),
if $l(y)=l(x)$ and a finite sequence $u$ exist such that
$x\sim_{w(i,n)} u$ and $\tilde F_{[i]_3}(y)\subseteq u$.

We can now define the recursive predicate which is needed to specify Template~2.
\begin{equation}\label{main-rel-1}
B(i,\sigma)\Longleftrightarrow
\sum\limits_{z}R_{[i]_2}(z)\le 2^{-(\sigma_1+3)},
\end{equation}
where $R_{[i]_2}$ denotes the frame of the flow through the elementary 
network $q_{[i]_2}$, and the sum is taken
over all $z$ of length $l(\sigma_2)$, which are $i$-discarded by an extra edge
$\sigma\in G^n_{[i]_1}(i)$.
Here in the exponent degree we identify the finite sequence $\sigma_2$ and its
order number.

Any task $i$ relates to the two networks: to its base $q_{[i]_1}$ and to its 
target $q_{[i]_2}$.
The goal of the task $i=\langle x_1,x_2,x_3\rangle$ is to provide conditions under
which the operator $F_{x_3}$ could not transform any infinite sequence 
from the support set of the semimeasure $P_{x_1}$ to a sequence from 
the support set of the semimeasure $P_{x_2}$.
A competing requirement is that these semimeasures should be nontrivial, i.e.
there should be $\bar P_{x_1}(\Omega)>0$ and $\bar P_{x_2}(\Omega)>0$.

{\it Construction of the networks $q_m$.}

We define the networks $q_m$ for $m=1,2,\dots$ using mathematical induction on 
steps $n=1,2,\dots$. 

Define $s_m(\lambda)=0$ and $G_m^0=\emptyset$ for all $m$.

Let $n\ge 1$ and for every $m$, the sets $G_m^{n-1}$ and 
the values $q_m(\sigma)$ be defined for all $\sigma\in G^{n-1}_m$, 
$s_m(x)$ be defined for all $x$ such that $l(x)\le n-1$ and
$q_m(\sigma)=\frac{1}{2}(1-s_m(\sigma_1)$ for each $\sigma$ of unit length such that
$l(\sigma_2)=n$.
                                     
At any step $n$ of induction execute the task $i=p(n)$:

1) denote $m=[i]_1$ and execute step $n$ of Template 2 with the predicate 
(\ref{main-rel-1}), where $s$, $q$ and $G^n$ are replaced with $s_m$, $q_m$ and $G^n_m$;

2) denote $m'=[i]_2$ and define $G^n_{m'}=G^{n-1}_{m'}$ and $s_{m'}(x)=1$ for each 
$x$ of length $n$, which is $i$-discarded at step $n$ by some edge 
$\sigma\in G^n_m(i)$ such that $l(\sigma_2)=n$ 
and define $s_{m'}(x)=0$ for all other $x$, $l(x)=n$;

3) for any $m\not=[i]_1$ and $m\not=[i]_2$ define $G^n_m=G^{n-1}_m$
and $s_m(x)=0$ for each $x$ such that $l(x)=n$;

4) After all, define
$
q_m(\sigma)=\frac{1}{2}(1-s_m(\sigma_1))
$
for each $m$ and for each $\sigma$ of unit length such that $l(\sigma_1)=n$.

This concludes the description of the induction step.

By Lemma~\ref{dup-1} $q_m(\sigma)=q_m(\sigma')$ for any
$\sigma,\sigma'\in G_m$ such that $\sigma\sim\sigma'$.

Let $P_m$ be the $q_m$-flow.

By Lemma~\ref{contin-1} the semimeasure $P_m$ is continuous for each $m$, since
the number $w(i)$ separates $G_m$ for each $i$.

The support set of any semimeasure $P_m$ is equal to
$$
E_{P_m}=\{\omega\in\Omega:\forall n(P_m(\omega^n)\not=0)\}.
$$

The following lemma is similar to Lemma~\ref{nontriv-1a} but its proof
has some new details.
\begin{lemma} \label{nontriv-1}
$\bar P_m({\bf 1})>0$ for each $m$.
\end{lemma}
{\it Proof.} Let us estimate from below the value of $\bar P_m(\Omega)$. Let
$R_m$ be the frame of the network $q_m$. Define
$$
S_{m,n}=\sum\limits_{u:l(u)=n}R_m(u)-
\sum\limits_{\sigma:\sigma\in G_m,l(\sigma_2)=n}q_m(\sigma)R_m(\sigma_1).
$$
By definition of the flow delay function,
\begin{eqnarray}
\sum\limits_{u:l(u)=n+1}R_m(u)=\sum\limits_{u:l(u)=n}(1-s_m(u))R_m(u)+
\label{RR-1a-1}
\\
\sum\limits_{\sigma:\sigma\in G_m,l(\sigma_2)=n+1}q_m(\sigma)R_m(\sigma_1).
\label{RR-2a-1}
\end{eqnarray}

Let $m$ be a base of the task $p(n)$ at step $n$, i.e., $m=[p(n)]_1$.

Let us first consider the case $w(p(n),n)<n$.
In this case the proof of this lemma coincides with the corresponding
part of the proof of Lemma~\ref{nontriv-1a},
where the delay function $s$ is replaced with the delay function $s_m$.
We pass this part of the proof and obtain $S_{m,n+1}\ge S_{m,n}$.

Consider the case $w(p(n),n)=n$. As in the proof of Lemma~\ref{nontriv-1a}
$\sum\limits_{u:l(u)=n}R_m(u)\le 1$ and
$$
\sum\limits_{u:l(u)=n}s_m(u)R_m(u)\le\rho(n)=1/(n+3)^2.
$$
Combining this inequality with (\ref{RR-1a-1})--(\ref{RR-2a-1}), we obtain
$S_{m,n+1}\ge S_{m,n-1}/(n+3)^2$.

Let $m$ be a target of the task $p(n)$ on step $n$, i.e., $m=[p(n)]_2$. Then
\begin{eqnarray}
\sum\limits_{l(u)=n}s_m(u)R_m(u)=\sum\limits_{u\in D} R_m(u)\le
\sum\limits_{\sigma\in G_m,l(\sigma_2)=n}2^{-(\sigma_1+3)},
\end{eqnarray}
where $D$ is the set of all $u$ of length $n$, which are
$p(n)$-discarded by sequences $\sigma_2$, where $\sigma\in G_{[p(n)]_1}$ 
and $l(\sigma_2)=n$.

Recall that in the exponent, we identify the finite sequence $\sigma_1$ and its
order number. Therefore,
$$
S_{m,n+1}\ge S_{m,n}-\sum\limits_{\sigma\in G_{[p(n)]_1},l(\sigma_2)=n}2^{-(\sigma_1+3)}.
$$
If $m$ is neither base no target of the task $p(n)$ at step $n$
(i.e. $[p(n)]_1\not=m$ and $[p(n)]_2\not=m$), then $s_m(u)=0$
for each $u$ of length $n$ and there is no edge $\sigma\in G_m$,
such that $l(\sigma_2)=n$. Hence, $S_{m,n+1}=S_{m,n}$.

Using these bounds for $S_{m,n}$ and $S_{m,0}=1$, we obtain
$$
S_{m,n}\ge 1-\sum\limits_{i=1}^{\infty} (i+3)^{-2}-
\sum\limits_{x\in\Xi}2^{-(x+3)}\ge\frac{1}{2}
$$
for all $n$. Since $P_m\ge R_m$, we have
$$
\bar P_m(\Omega)=\inf\limits_{n}\sum\limits_{l(u)=n} P_m(u)\ge
\inf\limits_n S_{m,n}\ge\frac{1}{2}.
$$
Lemma is proved. $\Box$

\begin{lemma}\label{non-equv-1}
If $k\not=m$ then $F_t(E_{P_k})\cap E_{P_m}=\emptyset$ for all $t$.
\end{lemma}
{\it Proof.} Assume that an $\omega\in E_k$ exists such that
$F_t(\omega)\in E_m$ for some $t$. Consider the task $i=\langle k,m,t\rangle$.
For any $n$, let $D_n$ be the set of all $z$ of length $n$ that are
$i$-discarded by the finite sequence $\omega^n$. It follows from
continuity of $P_m$ that
$$
\lim\limits_{n\to\infty}\sum\limits_{z\in D_n}P_m(z)\le
\lim\limits_{n\to\infty}2^{w(i,n)}P_m(\tilde F_t(\omega^n))=0.
$$
Besides, $P_k(\omega^n)\not=0$ for all $n$. From here it is easy to see
that for each $n$ the sequence $\omega^n$ has an $i$-extension
($\omega$ itself is suitable as such extension).

By Lemma~\ref{exten-1} an edge $\sigma\in G_k(i)$ will be drawn on some
step $n>w(i)$ such that $l(\sigma_2)=n$,
$\sigma_2\subset\omega$ and the sequence $(\tilde F_t(\omega))^n$ is $i$-discarded
by the sequence $\sigma_2$. Since $w(i,n)=w(i)$, no extra edge
$\sigma'$ such that $\sigma'_1\subset (\tilde F_t(\omega))^n\subset\sigma'_2$ can be
drawn at a step $n'=l(\sigma'_2)>n$, and therefore, no extra portion of the flow
can go through the vertex $(\tilde F_t(\omega))^n$.
Then $q_m(\big((\tilde F_t(\omega))^n,(\tilde F_t(\omega))^{n+1})\big)=0$ 
that implies $F_t(\omega)\not\in E_m$. The resulting contradiction proves the lemma.
$\Box$

\begin{theorem}\label{nucl-atom-1}
The set of all atoms of $\Upsilon$ is countable.
\end{theorem}
{\it Proof.}
Let ${\bf p}_m=[\bar E_{P_m}]$. By Lemma~\ref{nontriv-1} $P_m({\bf p}_m)>0$,
then ${\bf p}_m\not ={\bf 0}$. By Lemma~\ref{non-equv-1} for $k\not = m$,
any $\alpha\in E_{P_k}$ and $\beta\in E_{P_m}$ do not reduce to each other. Therefore,
${\bf p}_k\cap{\bf p}_m={\bf 0}$. Define ${\bf d}_m=i_{P_m}({\bf p}_m))$.
Since $\bar P_m(i_{P_m}({\bf p_m}))=\bar P_m({\bf p_m})$, we have
${\bf d}_m\not ={\bf 0}$.
From ${\bf d}_m\subseteq{\bf p}_m$ the equality
${\bf d}_k\cap{\bf d}_m={\bf 0}$ follows for $k\not = m$.

Assume that ${\bf d}_m={\bf a}\cup{\bf b}$, where ${\bf a}\not ={\bf 0}$,
${\bf b}\not ={\bf 0}$ and ${\bf a}\cap\bf b={\bf 0}$.
Then by Corollary~\ref{Rad-Nic} 
$\bar P_m({\bf a})>0$ and $\bar P_m({\bf b})>0$, which contradicts
Corollary~\ref{atom-el-1}.
This contradiction proves that ${\bf d}_m$ is an atom for each $m$.
Theorem is proved. $\Box$

Theorems~\ref{infini-div-1} and~\ref{nucl-atom-1} imply the main result
of this paper on decomposition of the maximal element of LV-algebra.
\begin{theorem} \label{nucl-ato-1}
It holds ${\bf 1}=\cup_{i = 1}^{\infty}{\bf a}_i\cup{\bf d}$,
where ${\bf a}_1, {\bf a}_2, \dots$ is the infinite sequence of all atoms and
${\bf d}$ is the non-zero infinitely divisible element.
\end{theorem}

\subsection{Decomposition of the hyperimmune LV-degree into atoms}\label{hyperimmune-2}

Let $\bf h$ be the element of $\Upsilon$ defined by the collection of
all hyperimmune sequences. We call this element hyperimmune LV-degree.  
Rumyantsev and Shen~\cite{RuS2014} proved that
hyperimmune sequences can be generated by some probabilistic machine
with positive probability. From this ${\bf h}\not={\bf 0}$ follows.
In this section we present a decomposition of $\bf h$ into a union of
the infinite sequence of atoms and of the infinitely divisible element.

An infinite subset $A\subseteq\cal N$ is called hyperimmune if there is no computable
function $f$ such that $f(i)\ge z_i$ for every $i$, where $z_1<z_2<\dots$ be all elements
of the set $A$ arranged in ascending order.
Let $a=a_1a_2\dots$ be the characteristic (binary) sequence of the set $A$, i.e.,
$a_i=1$ if and only if $i\in A$ for every $i$. We call $a$ hyperimmune sequence.
We will study LV-degrees generated by Turing degrees of hyperimmune sequences.

An infinite binary sequence $\alpha$ is called sparse if it contains infinitely
many ones and there is no computable
total function $f$ such that for each $k$ the prefix of $\alpha$ of length $f(k)$
contains at least $k$ ones.

\begin{proposition}\label{sparse-hyper-1}
A set $A$ is hyperimmune if and only if its characteristic sequence is sparse.
\end{proposition}
{\it Proof.} Assume that a set $A$ is not hyperimmune. Then $f(i)\ge z_i$, where
$z_1<z_2<\dots$ be all elements of the set $A$ arranged in ascending order.
It holds $a_{z_i}=1$ for all $i$ and $a_j=0$ for each $j\not\in A$. Since
$a_{z_i}=1$ for each $i$, the prefix of $a$ of length $f(i)$ contains at least
$i$ ones for each $i$, i.e., the sequence $a$ is not sparse.

On other side, assume that the characteristic sequence $a$ of the set $A$ is not
sparse. Then there is a computable function $f$ such that for each $k$ the prefix
of $a$ of length $f(k)$ contains at least $k$ ones. Since $a_{z_i}=1$ for each $i$,
$f(k)\ge z_k$, i.e., the set $A$ is not hyperimmune.
$\Box$

More information about the hyperimmune LV-degrees can be found in 
Holzl and Porter~\cite{HoP2021}, Proposition~4.15. 

The following theorems~\ref{nucl-4-1},~\ref{nucl-4-2}, and~\ref{nucl-4}
present a decomposition of the hyperimmune degree into the union of a countable 
sequence of atoms and a non-zero infinitely divisible element.  

\begin{theorem}\label{nucl-4-1}
There exists an infinite sequence $ {\bf h}_1, {\bf h}_2, \dots$
of atoms defined by collections of hyperimmune sequences.
\end{theorem}
{\it Proof.}
We modify Template 2 for the case of two recursive predicates
$B_1(j,\sigma)$ and $B_2(j,\sigma)$.

We call $i=p(l(x))$ the atoms difference task if $i$ is even, $i=2j$,
and we call $i=p(l(x))$ the sparsity task if $i$ is odd, $i=2j+1$.

We say that a finite sequence $x$ of length $n$ is $j$-discarded by a sequence
$y$ (or by an edge $\sigma\in G^n_{[j]_1}$ such that $\sigma_2=y$),
if $l(y)=l(x)=n$ and a finite sequence $u$ exist such that
$x\sim_{w(i,n)} u$ and $\tilde F_{[j]_3}(y)\subseteq u$.

Let us define the first predicate which have to provide the difference
between atoms:
\begin{equation}\label{rel-1-1}
B_1(j,\sigma)\Longleftrightarrow 
\sum\limits_{z}R_{[j]_2}(z)\le 2^{-\sigma_1+3},
\end{equation}
where by $R_{[j]_2}$ we denote the frame of the flow through
the elementary network $q_{[j]_2}$ defined on steps $<n$,
and the sum is taken over all $z$ of length $l(\sigma_2)$, which are $j$-discarded
by the sequence $\sigma_2$. Here, in the exponent, we identify the sequence
$\sigma_1$ and its number.

Let $\phi_j$ be a computable sequence of all partial recursive functions
such that for any partial recursive function $f$ there exist infinitely many $j$
such that $\phi_j=f$, $\phi^n_i(x)$ is a result of computation in $n$ steps
(see Section~\ref{prelimin-1}).

Define the second predicate which have to provide the sparsity of sequences from
the support set of the $q_{[j]_1}$-flow:
\begin{eqnarray*}
B_2(j,\sigma)\Longleftrightarrow 
\sigma_2=\sigma_110^{l(\sigma_2)-l(\sigma_1)-1}\&
l(\sigma_2)\ge\phi^{l(\sigma_2)}_{[j]_1}(l(\sigma_1)+2).
\end{eqnarray*}

{\it Construction of the network $q_m$.}

The induction hypothesis is the same as for step $n$ of Template 2.

At any step $n$ of induction we execute the task $i=p(n)$. This means that

1) Let $i=2j$. In this case do the following:

1.1) denote $m=[j]_1$ and execute step $n$ of Template~2 with the
predicate $B_1(j,\sigma)$ to define the set $G^n_m$, the values $q_m(\sigma)$
for $\sigma\in G^n_m$ such that $l(\sigma_2)=n$, and the values of the
flow delay function $s_m(x)$ for all $x$ of length $n$;

1.2) denote $m'=[j]_2$ and define $G^n_{m'}=G^{n-1}_{m'}$,
$s_{m'}(x)=1$ for each $x$, which is $j$-discarded on step $n$
by some edge $\sigma\in G^n_{[j]_1}(j)$; define $s_{m'}(x)=0$ for all other
$x$ such that $l(x)=n$;

1.3) for each $m$ such that $m\not=[j]_1$ and $m\not=[j]_2$ define 
$G^n_m=G^{n-1}_m$ and $s_m(x)=0$ for every $x$ of length $n$.

2) Let $i=2j+1$. In this case do the following:

2.1) denote $m=[j]_1$ and execute step $n$ of Template 2 with the
predicate $B_2(j,\sigma)$ to define the set $G^n_m$, the values $q_m(\sigma)$
for $\sigma\in G^n_m$ such that $l(\sigma_2)=n$, and the values of the
flow delay function $s_m(x)$ for all $x$ of length $n$;

2.2) for each $m\not=[j]_1$ define $G^n_m=G^{n-1}_m$ and
$s_m(x)=0$ for every $x$ of length $n$;

3) after all, for every $m$ define 
$q_m(\sigma)=\frac{1}{2}(1-s_m(\sigma_1))$
for each $\sigma$ of unit length such that $l(\sigma_1)=n$.

This concludes the description of the induction step.

By Lemma~\ref{dup-1} $q_m(\sigma)=q_m(\sigma')$ for any
$\sigma,\sigma'\in G_m$ such that $\sigma\sim_{w(i,n)}\sigma'$.


Let $P_m$ be the $q_m$-flow. Define ${\bf p}_m=[\bar E_{P_m}]$.
By Lemma~\ref{nontriv-1} $P_m({\bf p}_m)>0$, then ${\bf p}_m\not ={\bf 0}$.

Define ${\bf h}_m=i_{P_m}({\bf p}_m)$ for each $m$.
${\bf h}_m\not ={\bf 0}$, since $\bar P(i_P({\bf p_m}))=\bar P({\bf p_m})$.
The LV-degree ${\bf h}_m$ is an atom of $\Upsilon$ for each $m$,
since we use Template 2 for its definition.

By Lemma~\ref{non-equv-1}, for $k\not=m$,
any $\alpha\in E_{P_k}$ and $\beta\in E_{P_m}$ do not Turing reducible to
each other. Therefore, ${\bf p}_k\cap{\bf p}_m={\bf 0}$.
Since ${\bf h}_m\subseteq{\bf p}_m$, we obtain ${\bf h}_k\cap{\bf h}_m={\bf 0}$
for $k\not=m$.
$\Box$

The rest of the proof of Theorem~\ref{nucl-4-1} is presented in the following lemma.
\begin{lemma} \label{nontriv-1b-3}
Any infinite sequence $\omega$ from the support set of the semimeasure $P_m$ is sparse.
\end{lemma}
{\it Proof.} Let $m$ be given.
We should prove that for any infinite sequence $\omega$ from the support set of
the semimeasure $P_m$ and for any total computable function $f$, there are infinitely
many $k$ such that the prefix of $\omega$ of length $f(k)$ contains less than $k$ ones.

For any computable function $f$ there are infinitely many odd $i=2j+1$
such that $f=\phi_j$. Since $f$ is total, each prefix of any $\omega\in E_{P_m}$ has
an $j$-extension. By Lemma~\ref{exten-1} $\sigma_1\subset\sigma_2\subset\omega$
for infinitely many extra edges $\sigma$ such that $f(l(\sigma_1)+2)\le l(\sigma_2)$.
Since number of ones in $\sigma_2=\sigma_110^{l(\sigma_2)-l(\sigma_1)-1}$ is
less or equal to $l(\sigma_1)+1$ and $f(l(\sigma_1)+2)\le l(\sigma_2)$, the prefix
of $\omega$ of length $f(l(\sigma_1)+2)$ contains less than $l(\sigma_1)+2$ ones.

Since at least one 1 is added to $\sigma_1$ at infinitely many steps,
the sequence $\omega$ contains infinitely many 1s. Hence, each $\omega$
from the support set of semimeasure $P_m$ is sparse.
$\Box$

\begin{theorem}\label{nucl-4-2}
There exists an infinitely divisible element defined by a collection of
the hyperimmune sequences.
\end{theorem}
The proof is similar to the proof of Theorem~\ref{nucl-4-1}, where
the recursive predicate $B_1$ is replaced with (\ref{rel-2})
and the Template 1 is used.

Theorems~\ref{nucl-4-1} and~\ref{nucl-4-2} imply the main result of
Section~\ref{hyperimmune-2}.
\begin{theorem} \label{nucl-4}
The decomposition ${\bf h}=\cup_{i=1}^{\infty}{\bf h}_i\cup{\bf e}$
of hyperimmune LV-degree takes place, where $ {\bf h}_1, {\bf h}_2,\dots$ are
infinite sequence of atoms and ${\bf e}$ is the infinitely divisible element
defined by collections of hyperimmune sequences.
\end{theorem}

It should be interesting to extend the result of Theorem~\ref{nucl-4}
to other specific LV-degrees.
A careful analysis of the relationship between LV-degrees and Turing degrees is
given in the review by Holzl and Porter~\cite{HoP2021}. We have proved that
some of LV-degrees can be generated using Template 2 and, so,
the decomposition of type (\ref{main-decomposition-2}) takes place for these LV-degrees.
For example, this is the hyperimmune degree. Holzl and Porter~\cite{HoP2021}
result on DNC (diagonally non-computable)
degree can be extended to obtain the decomposition like (\ref{main-decomposition-2})
for this degree.\footnote{An infinite binary sequence $\omega$ has DNC degree
if and only if there is some function $f$ such that $f\le_T\omega$
and $f(i)\not=\phi_i(i)$ for all $i$.}

An open problem arises can we obtain decompositions of type (\ref{main-decomposition-2})
of the LV-degrees considered in~\cite{HoP2021} among which there are degrees defined
by the collection of 1-generic sequences, degrees defined by the collection of
generalized low sequences, and those collections corresponding to various notions
of effective randomness. Author does not know whether it is possible to apply
the technics of Templates 1 and 2 for the construction of LV-degrees of
1-generic sequences.

\end{document}